\documentclass[prd,showpacs,amsmath,showkeys,floatfix,amssymb, preprintnumbers, nofootinbib, superscriptaddress,reprint]{revtex4-1}
\usepackage{graphicx,color}
\usepackage{amsmath,dsfont}
\usepackage{amssymb}
\usepackage{booktabs}
\usepackage{mathrsfs}
\usepackage{cases}
\usepackage[capitalise]{cleveref}
\usepackage{chngcntr}
\usepackage{floatrow}
\def\simle{\mathrel{\rlap{\raise 0.511ex \hbox{$<$}}{\lower 0.511ex \hbox{$\sim$}}}}
\def\simge{\mathrel{ \rlap{\raise 0.511ex \hbox{$>$}}{\lower 0.511ex \hbox{$\sim$}}}}

\newcommand \beq{\begin{flalign}}
\newcommand \eeq{\end{flalign}} 
\def\simle{\mathrel{\rlap{\raise 0.511ex \hbox{$<$}}{\lower 0.511ex 
\hbox{$\sim$}}}}
\def\simge{\mathrel{ \rlap{\raise 0.511ex 
\hbox{$>$}}{\lower 0.511ex \hbox{$\sim$}}}}

\newcommand\lb{\left(}
\newcommand\rb{\right)}
\newcommand{\lsb}{\left[ }
\newcommand{\rsb}{\right]}
\newcommand{\llb}{\left\lbrace}
\newcommand{\rrb}{\right\rbrace}
\newcommand{\lan}{\langle}
\newcommand{\ran}{\rangle}
\newcommand{\dsum}{\displaystyle\sum}

\newcommand\pt{\partial}
\newcommand*\diff{\mathop{}\!\mathrm{d}}
\newcommand*\Diff[1]{\mathop{}\!\mathrm{d^#1}}
\newcommand{\Tr}{{\rm Tr}}

\newcommand{\nb}{\nonumber}

\renewcommand{\Re}{\hbox{Re}}

\begin{document}

\title{Confinement, Holonomy and Correlated Instanton-Dyon Ensemble I: SU(2) Yang-Mills Theory}

\author{Miguel Angel Lopez-Ruiz}
\email{malopezr@indiana.edu}
\affiliation{ Physics Department and Center for Exploration of Energy and Matter,
Indiana University, 2401 N Milo B. Sampson Lane, Bloomington, IN 47408, USA.}

\author{Yin Jiang}
\email{y.jiang@thphys.uni-heidelberg.de}
\affiliation{ Physics Department and Center for Exploration of Energy and Matter,
Indiana University, 2401 N Milo B. Sampson Lane, Bloomington, IN 47408, USA.}
\affiliation{School of Physics and Nuclear Energy Engineering, Beihang University, Beijing 100191, China.}

\author{Jinfeng Liao}
\email{liaoji@indiana.edu}
\affiliation{ Physics Department and Center for Exploration of Energy and Matter,
Indiana University, 2401 N Milo B. Sampson Lane, Bloomington, IN 47408, USA.}

\date{\today}

\begin{abstract}
The mechanism of confinement in Yang-Mills theories remains a  challenge to our understanding of nonperturbative gauge dynamics. While it is widely perceived that confinement may arise from chromo-magnetically charged gauge configurations with nontrivial topology, it is not clear what types of configurations could do that and how, in pure Yang-Mills and QCD-like (non-supersymmetric) theories.  Recently, a promising approach has emerged, based on statistical ensembles of dyons/anti-dyons that are constituents of instanton/anti-instanton solutions with nontrivial holonomy where the holonomy plays a vital role as an effective ``Higgsing'' mechanism. We report a thorough numerical investigation of the confinement dynamics in $SU(2)$ Yang-Mills theory by constructing such a  statistical ensemble of correlated instanton-dyons.  
\end{abstract}

\maketitle

\section{Introduction}

The Quantum Chromodynamics, or QCD, is established as the fundamental quantum field theory of strong nuclear force underlying all of nuclear physics. Despite its great success in describing an impressive variety of nuclear phenomena in nature, a key aspect of QCD remains mysterious and poses a great challenge to our understanding. While the theory has quarks and gluons as its basic degrees of freedom in the Lagrangian, the colored quarks and gluons are absent from the observed physical spectrum in which the various color-singlet hadronic states emerge instead. This phenomenon, often referred to with the broad term ``confinement'' (--- see recent review in e.g.~\cite{Greensite:2011zz}), occurs also in a wide variety of QCD-like theories and notably in pure Yang-Mills theories. The latter fact makes it obvious that confinement arises from the nonperturbative gauge dynamics in the gluonic sector. It was suggested long ago~\cite{'tHooft:1981ht,Nambu:1974zg,Mandelstam:1974pi}, based on analogy with superconductivity, that the confinement may arise from chromo-magnetically charged and topologically nontrivial gauge configurations, with the vacuum being a ``dual superconductor'' of such magnetic objects (--- see review in e.g.~\cite{Ripka:2003vv,Kondo:2014sta} ).  This scenario is highly appealing and widely perceived to be a likely mechanism for confinement. The idea was extensively explored via lattice simulations as well as has been concretely   shown to work in certain supersymmetric theories~\cite{Seiberg:1994rs,Seiberg:1994aj}. More recently, the idea of ``dual superconductor'' vacuum has been further advanced into a ``magnetic scenario'' for the hot plasma phase in the temperature regime above but near the transition temperature $T_c$~\cite{Liao:2006ry}: indeed, if there is a magnetic superconductor below $T_c$, there should be a ``precursor'', i.e., the normal phase of thermal magnetic plasma just above $T_c$. There are strong evidences from lattice simulations~\cite{Chernodub:2006gu,D'Alessandro:2007su,Bonati:2013bga} for this  scenario and most interestingly such a magnetic component is found to be crucial for explaining a number of key transport properties of the near-$T_c$ QCD plasma  as measured from heavy ion collision experiments~\cite{Liao:2008dk,Ratti:2008jz,Xu:2014tda} (--- see e.g.~\cite{Shuryak:2014zxa} for a recent review). Despite the various progress so far along this line of thought, it is nevertheless still unclear what types of configurations could drive confinement and how, in pure Yang-Mills and QCD-like (non-supersymmetric) theories. It shall be mentioned in passing that there exist a variety of interesting alternative ideas about possible mechanism and possible topological objects that may drive the confinement~\cite{Zhitnitsky:2008ha,Zhitnitsky:2006sr,Poppitz:2012sw,Poppitz:2011wy,Kondo:2016ywd}.  
     
Let us elaborate a bit on the difficulty to identify the relevant topological configurations for confinement in these theories. Conventional instantons (as well as their finite-temperature counterpart, the calorons)~\cite{Belavin:1975fg,Harrington:1978ve} are well known topological objects  and studied in detail (--- see reviews in e.g.~\cite{Schafer:1996wv,Diakonov:2002fq}). However, these conventional  instantons/calorons only have {\em trivial holonomy}, that is, trivial Polyakov loop at spatial infinity $|\vec x| \to \infty$ (--- see \ \cref{AP:Hlmy,AP:HolPol} for a detailed discussion about holonomy), which is in sharp contrast to the confining vacuum where the holonomy is maximally nontrivial. Furthermore they are color neutral with no chromo-magnetic charge. Therefore, such conventional instantons/calorons can not be responsible for confinement. Another natural candidate would be 't Hoof-Polyakov type of magnetic monopole (--- see e.g. review in ~\cite{tHooft:1999cgx}). However, a crucial difference of pure Yang-Mills or QCD (as compared with e.g. George-Glashow model or Seiberg-Witten therory) is that they do {\em not} have adjoint scalar fields which would provide the natural ``Higgsing''  on the spatial boundary, thus allowing the construction of magnetic monopole solutions. Because of this difficulty, one usually has to rely upon certain gauge-fixing procedure to manifest the monopoles in the pure Yang-Mills or QCD cases. 
 
As it turns out,  both difficulties are resolved recently in a crucial new development: the construction of {\em caloron solutions with nontrivial holonomy and nontrivial topology} to the classical Yang-Mills equations, known as the KvBLL calorons~\cite{Kraan:1998pm,Kraan:1998sn,Lee:1998bb} (---  see \cref{AP:KvBLL} for a detailed discussion). First of all, such new calorons by construction acquire the necessary sensitivity to holonomy and thus are able to play a role in the confinement dynamics. Even more nontrivially, such a caloron is made of $N_c$ different {\em constituent dyons} (for $SU(N_c)$ theory) which are chrome-magnetically charged and whose properties critically depend upon the holonomy. In these solutions, the nontrivial holonomy provides the needed nontrivial boundary constraints (which  would have to come from the adjoint scalar fields in ``Higgsed'' theories). Owing to such important new features, the KvBLL calorons with their dyon constituents provide the unique topological configurations that could potentially account for the nonperturbative dynamics underlying confinement.  
  
Based on the KvBLL solutions, a promising approach has emerged for understanding confinement in a statistical ensemble of dyons/anti-dyons arising from the constituents of the KvBLL calorons. Early works on an uncorrelated ensemble of these objects --- to be called {\em instanton-dyons} (following recent literature) from now on in the present paper --- already indicated that their contributions (alone) to the holonomy potential tends to push the system toward confining holonomy~\cite{Diakonov:2007nv,Diakonov:2009jq,Gerhold:2006sk}. However it was later found that such contributions would not be enough to overcome the one-loop perturbative contributions to the holonomy potential (which favors the trivial holonomy --- see detailed discussions in \cref{AP:HolPol}). It was later found by Shuryak and collaborators~\cite{Shuryak:2013tka,Faccioli:2013ja,Larsen:2014yya,Larsen:2015vaa,Larsen:2015tso} that an effective dyon--anti-dyon interaction with a short-range repulsive core (or in a broader term,  a strong dyon--anti-dyon short-range correlation) appears necessary to enforce the confining holonomy at low temperature.  Effective model for a Coulomb plasma of such dyons/anti-dyons was also constructed and shown to give a reasonable qualitative description of the low temperature $T<T_c$ properties of Yang-Mills theories~\cite{Liu:2015ufa,Liu:2015jsa}. 

In this paper, we report a thorough numerical investigation of the confinement dynamics in $SU(2)$ Yang-Mills theory by constructing such a  statistical ensemble of correlated instanton-dyons. We present high precision results for the temperature dependence of the holonomy potential, the order parameter for confinement transition, the dyon ensembe properties (densities and density-density correlations), as well as the temporal and spatial Wilson loops. In particular, we study the influence of the finite volume effect, the dyon--anti-dyon correlations as well as the screening mass on the confinement dynamics. Some of these results are considerably improved as compared with previous studies and many new results have not been previously reported. 

The rest of the paper is organized as follows. In Section 2, we present the detailed construction of the correlated instanton-dyon ensemble as well as the numerical procedures. Our main results about the confinement dynamics in such an ensemble are reported in Section 3. The Section 4 then focuses on examining the consequences of the key parameters in the ensemble construction. Finally, we conclude the study in Section 5. In addition a few Appendices are included to explain some ``background'' information in detail and to make the paper more self-contained for the convenience of readers.

\section{Construction of Correlated Instanton-Dyon Ensemble}

\subsection{The Partition Function} 

The construction of the partition function of the dyon ensemble begins with rewriting the one-loop quantum weight of a single KvBLL caloron in the limit of large dyon separation \cref{Eq:Z_KvBLL3} (--- see  \cref{AP:KvBLL,AP:KvBLL_QW} for more details), namely the contribution of a pair of $L$ and $M$ dyons, in the following way: 

\begin{equation}
 \mathcal{Z} = e^{-VP(\nu)/T}\int \lb\Diff{3}r_L \,f_L \rb \lb\Diff{3}r_M\, f_M \rb T^6 \det(\hat{G}) \, .
\end{equation} 
Here, the fugacities per dyon species are introduced as $f_M=\Gamma S^2 e^{-\nu S}\nu^{\frac{8\nu}{3}-1}$ and $f_L=\Gamma S^2 e^{-\bar{\nu} S}\bar{\nu}^{\frac{8\bar{\nu}}{3}-1}$. The $P(\nu)$ is the famous Gross-Pisarski-Yaffe result for perturbative contributions while the $\det{\hat{G}}$ is the contribution of one-loop quantum fluctuation around one KvBLL caloron. Extending this result to arbitrary number of $L$ and $M$ dyons requires the inclusion of the appropriate metric of the moduli space which, as of today, its explicit form has not yet been found. Nevertheless, Diakonov and Petrov \cite{Diakonov:2007nv} proposed an approximate metric by merging that of a neutral cluster of dyons of different kind, namely, an $L$-$M$ pair, with that of dyons of the same kind (originally proposed in \cite{Gibbons:1979xn}). Therefore, the full measure is approximated by the square of the determinant of a symmetric matrix $G$ like $\sqrt{\det(g)}\approx \det(G)$. Despite not being an exact solution, it possesses the interesting property that in the limit of $K$ well separated $L$-$M$ pairs, the measure factorizes into $\det(G)=\det(\hat{G})^K$, i.e., as the product of $K$ individual KvBLL caloron measures.  It is thus straightforward to see that for a single $L$-$M$ pair, the $G$ matrix reduces to $\hat{G}$ (\cref{Eq:G_KvBLL}). In the $SU(2)$ case, when the number of $L$ and $M$ dyons are  $N_L$ and $N_M$ respectively, the dimension of this matrix $G$ is $(N_L+N_M)\times (N_L + N_M)$ and its components are given by: 

\begin{flalign}
\nonumber         
G_{mi,nj} &=\delta_{mn}\delta_{ij}\lb 4\pi\nu_m - \sum_{k\neq i}\frac{2}{T|\vec{r}_{m_i}-\vec{r}_{m_k}|}\right. \\
\nonumber
&\left. + \sum_{k}\left.\frac{2}{T|\vec{r}_{m_i}-\vec{r}_{l_k}|}\right|_{m\neq l}\rb + \left.\frac{2\delta_{mn}}{T|\vec{r}_{m_i}-\vec{r}_{n_j}|}\right|_{i\neq j}\\
&   - \left.\frac{2}{T|\vec{r}_{m_i}-\vec{r}_{n_j}|}\right|_{m\neq n},
\label{Eq:G_def}
\end{flalign}  
where $\vec{r}_{m_i}$ is the position vector of the $i^{\text{th}}$ dyon of kind $m$ (either $L$ or $M$). Furthermore, it should be clear that similar results can be obtained for anti-dyons.

As pointed out in \cite{Diakonov:2009jq}, dyon--anti-dyon configurations are not saddle points of the Yang-Mills action. The inclusion of anti-self-dual fields in the ensemble is done by factorizing the measure into uncorrelated parts $\det(G_D)\det(G_{\bar{D}})$ ($D$ for dyons and $\bar{D}$ for anti-dyons) times a correlated contribution $e^{-V_{D\bar{D}}}$, where $V_{D\bar{D}}$ is the action corresponding to dyon--anti-dyon interactions. Classical interactions between dyon--anti-dyon of the same kind was recently introduced in a gradient flow study in \cite{Larsen:2014yya}. Using the parametrization found in \cite{Larsen:2015vaa}, the potential takes the following form

\begin{flalign}
\nb
V_{L\bar{L}} = -2\bar{\nu} S\lb\frac{1}{\zeta_L} -1.632 e^{-0.704\zeta_L}\rb,\\
\nb
V_{M\bar{M}} = -2\nu S\lb\frac{1}{\zeta_M} -1.632 e^{-0.704\zeta_M}\rb, \\
\zeta_j = 2\pi\nu_j T r_{j\bar{j}} ,
\label{Eq:VDD1}
\end{flalign} 
for $\zeta_j > \zeta_j^c$ and $r_{j\bar{j}}=|\vec{r}_j-\vec{r}_{\bar{j}}|$ . Below the limit $\zeta_j < \zeta_j^c$, the interaction is repulsive and the proposed core potential for this region is given by

\begin{equation}
V^C_{j\bar{j}} = \frac{\nu_j V_c}{1+e^{(\zeta_j - \zeta_j^c)}} \,\, ,
\label{Eq:VDD_Core}
\end{equation} 
where $V_c$ and $\zeta_j^c$ are the two key parameters that quantify the strength and range of the repulsive correlations between dyon--anti-dyon pairs. 

Other interactions that have to be accounted for include the long-range forces due to the Abelian electric and magnetic charges and the nonlinear terms in the field strength tensor $F_{\mu\nu}$, given by

\begin{equation}
V_{ij}=\frac{S}{2\pi T r_{ij}}(e_ie_j + m_im_j -2h_ih_j),
\label{Eq:VDD2}
\end{equation} 
where $e_j$ and $m_j$ are the electric and magnetic charges (--- see \cref{Tab:Dyon_Prop}) and $h_j=1$ for the $M$-type (anti)dyons while $h_j=-1$ for the $L$-type ones. As expected, this gives exactly cancelled classical interaction between the $L$ and $M$ dyons (as well as $\bar{L}$ and $\bar{M}$) that together make a KvBLL caloron (owing to their BPS nature). On the other hand, there is repulsive interaction  for the $L\bar{M}$ and $M\bar{L}$ pairs, while attractive interaction for the $L\bar{L}$ and $M\bar{M}$ pairs. 

In the construction of the ensemble, one has to sum over different number of (anti)dyons and also take into account the many-body screening effect which is introduced by means of a Debye screening mass $M_D$. In doing so, all Coulomb terms appearing in the partition function (including those in the $G$ matrices) are modified into  $r^{-1}\to r^{-1}e^{-M_Dr}$.  Combining \cref{Eq:VDD1,Eq:VDD_Core,Eq:VDD2}, the contribution from the inter-particle interactions to the action in the partition function is given by: 

\begin{widetext}
\begin{equation}
V_{D\bar{D}} = \llb \begin{array}{ll}
\vspace*{4mm}
-\dsum\limits_{j,\bar{j}} {2\nu_jS\lb\dfrac{1}{\zeta_j} - 1.632\, e^{-0.704 \zeta_j}\rb e^{-M_Dr_{j\bar{j}}}}& \text{if } \zeta_j>\zeta_j^c, \text{ for }L\bar{L},M\bar{M} \\
\vspace*{4mm}
\dsum\limits_{\substack{i>j\\j,\bar{j}}} V_{ij}^C & \text{if } \zeta_j<\zeta_j^c, \text{ for } \begin{array}{l}
LL, \bar{L}\bar{L}, MM, \bar{M}\bar{M},\\
L\bar{L},M\bar{M}
\end{array}\\
\vspace*{4mm}
\dsum\limits_{i,\bar{j}}\dfrac{S}{\pi Tr_{i\bar{j}}}e^{-M_D r_{i\bar{j}}} &\text{for }\bar{M}L, \bar{L}M\\
0 &\text{for } LM, \bar{L}\bar{M}.
\end{array} \right.
\end{equation} 
With all the above elements, one finally writes down the following  form for the full partition function of the dyon--anti-dyon ensemble:

\begin{align}
\nb
\mathcal{Z} &= e^{-VP(\nu)/T} \sum_{\substack{N_M,N_L, \\ N_{\bar{L}}, N_{\bar{M}}}} \frac{1}{N_L!N_M!N_{\bar{L}}!N_{\bar{M}}!} \int \prod_{l=1}^{N_L} f_L T^3\Diff3r_{L_l} \prod_{m=1}^{N_M} f_M T^3\Diff3r_{M_m} \\
& \times \prod_{\bar{l}=1}^{N_{\bar{L}}}f_{\bar{L}} T^3\Diff3r_{\bar{L}_{\bar{l}}} \prod_{\bar{m}=1}^{N_{\bar{M}}} f_{\bar{M}} T^3\Diff3r_{\bar{M}_{\bar{m}}} \det(G_D)\det(G_{\bar{D}}) \,e^{-V_{D\bar{D}}},
\label{Eq:Part_Func}
\end{align}
\end{widetext} 
where the factorial terms are needed to avoid duplicate counting of identical configurations with given numbers of dyons and anti-dyons. By requiring neutrality condition, i.e., equal number of dyons and anti-dyons of the same kind, the above expression can be further simplified into:

\begin{flalign}
\nb
\mathcal{Z} &=e^{-VP(\nu)/T} \sum_{N_L,N_M}\lsb\frac{(f_LVT^3)^{N_L}}{N_L!}\frac{(f_MVT^3)^{N_M}}{N_M!}\rsb^2 \\
&\times e^{-VT^3\mathcal{F}(T,\nu)},
\label{Eq:Part_Func2}
\end{flalign} 
where $V$ is the 3D volume available and 

\begin{flalign}
\nb
e^{-VT^3\mathcal{F}(T,\nu)}&\equiv \int\prod_{\substack{l,m, \bar{l},\bar{m}}}^{N_L,N_M} \frac{\Diff3 r_{L_l}}{V}\frac{\Diff3 r_{M_m}}{V}\frac{\Diff3 r_{\bar{L}_{\bar{l}}}}{V}\frac{\Diff3 r_{\bar{M}_{\bar{m}}}}{V} \\
\nb
& \times \exp\llb \log \lsb \det(G_D)\det(G_{\bar{D}})\rsb - V_{D\bar{D}} \rrb\\
\label{Eq:Part_Func3}
\end{flalign}
is obtained after performing integrals over all dyon positions. Finally, using Stirling's approximation $\log N!\approx N\log N - N + \log\sqrt{2\pi N}$ and defining the dimensionless dyon densities as $n_D=N_D/VT^3$, we rewrite $\mathcal{Z}$ as a sum of weights running over number of dyons as

\begin{flalign}
\nb
\mathcal{Z} &= \sum_{N_L,N_M} \exp     \lsb -VT^3 \lb \frac{4\pi^2}{3}\nu\bar{\nu} + 2n_L\log\lb\frac{n_L}{f_L}\rb   \right.\right.\\
\nb
&  + 2n_M\log\lb\frac{n_M}{f_M}\rb + 2(n_L+n_M) + \frac{\log\lb 4\pi^2 N_LN_M\rb}{VT^3}   \\
\nb
& \left.\left.  + \mathcal{F}(T,\nu) \rb \rsb\\
& \equiv   \sum_{N_L,N_M} \mathcal{Z}_{\text{LM}}.
\label{Eq:Part_Func4}
\end{flalign}

In this framework, there are three key parameters as theoretical inputs: the screening mass $M_D$, as well as the strength parameter $V_C$ and range parameter $\zeta^c_j$ for the dyon--anti-dyon interaction potential. In principle these parameters could be constrained by comparing relevant observables from the dyon ensemble with lattice simulations. Such quantitative comparison will be the goal of a forthcoming study, while the present paper focuses on qualitative question of demonstrating how the confinement is driven to occur in the correlated dyon ensemble.

\subsection{The Monte-Carlo Simulations}

Let us now discuss the details of the Monte-Carlo simulations to be used for evaluating the dyon ensemble partition function. Different from the implementation in \cite{Faccioli:2013ja,Larsen:2015vaa,Larsen:2015tso}, in our simulation we used a flat geometry, namely a box with periodic boundary conditions which shall be  a more ``realistic" approach  and a more direct way to compare the results with e.g.  lattice simulations.  

From \cref{Eq:Part_Func4}, it can be seen that all explicit dependence on the temperature $T$ can be absorbed by rescaling $rT\to r$, $VT^3\to V$, $M_D/T\to M_D$ and the free energy $F/T\equiv -\log\mathcal{Z}\to F$. Since this simplifies the calculations, all the simulations are done using such scaled dimensionless variables. In doing so, the temperature $T$ superficially disappears from the explicit simulations. However, the temperature dependence implicitly affects the system properties through the running coupling constant in the caloron action $S$, which at one loop level is given by (--- see \cref{AP:KvBLL_QW}) 

\begin{equation}
S(T)=\frac{8\pi^2}{g^2(T)}=\frac{22}{3}\log\lb\frac{T}{\Lambda}\rb,
\label{Eq:S_Inst}
\end{equation} 
where $\Lambda$ is the scale parameter in the regularization. Therefore, by varying $S$ as a parameter in the simulation, one is essentially varying the system temperature. It is straightforward to convert $S$ into $T/\Lambda$. To further put temperature in e.g. MeV unit, one would then have to make a physical choice for the value of $\Lambda$. For example, one may choose $\Lambda$ such that the critical temperature $T_c$ matches the lattice obtained value for $SU(2)$ Yang-Mills theory.  Once $T_c$ is fixed, one can then measure temperature $T$ in terms of $T/T_c$ (--- note this is equivalent to specifying the ratio $T_c/\Lambda$). 

\begin{figure}[t]
\centering
\includegraphics[width=\textwidth]{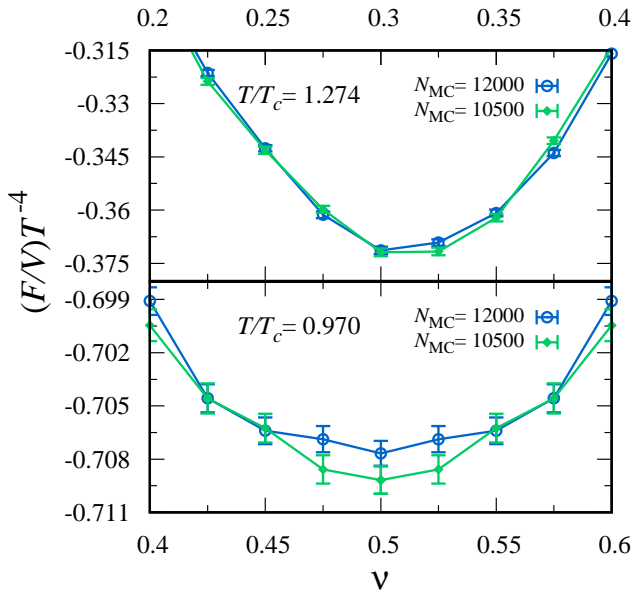}
\caption{Free energy density $F/V$ as a function of holonomy $\nu$ computed with different choices for the number $N_{\text{MC}}$ of the Monte Carlo configurations in the simulations,  in the confined (bottom) and deconfined (top) phases.}
\label{Fig:FreeEner_MCTest}
\end{figure}

The computation of all the observables are performed through Monte Carlo simulations using the well known Metropolis-Hastings algorithm. Each configuration is generated by first randomly varying the 3D positions of a single dyon or anti-dyon of each kind (and accounting for the periodic boundary conditions), then applying the acceptance algorithm, and moving to the next set of dyons/anti-dyons by repeating the same procedure. Once all positions have been swept, we then move to compute the observables with this new configuration and repeat all over again until the ensemble has been thermalized with enough statistics. It has been found that after about 2000 Monte Carlo configurations, the system is typically stabilized, after which the ensemble average would be calculated with the 10000 subsequent new configurations. The determined autocorrelation time was close to 5 configurations; therefore, the observables are averaged over 2000 configurations. On \cref{Fig:FreeEner_MCTest} we compare the free energy density calculated with a smaller number of Monte Carlo configurations for both confined and deconfined phases. The results are obviously consistent with each other and the small discrepancy between the two data sets is merely at a level of  approximately 0.54\% in the order parameter calculation $\left\langle L_\infty\right\rangle$. This  comparison clearly  demonstrates that with 12000 configurations one obtains very reliable results with rather small statistical uncertainty.

One technical issue in the Monte Carlo sampling process is about the measure factors $\det(G)$. As it is an approximation to the actual moduli space metric, it may happen that some of the eigenvalues of the $G$ matrices become negative thus violating the positive definiteness of the metric. This issue has been addressed in detail by \cite{Diakonov:2009jq,Liu:2015ufa,Bruckmann:2009nw}. To avoid configurations with negative eigenvalues in the simulations, the approach taken was to use the Metropolis-Hastings algorithm to reject such ``wrong'' configurations by assigning them a small statistical weight, i.e. if either $G_D$ or $G_{\bar{D}}$ has at least one negative eigenvalue, then the weight $\exp(\log\det(G))$ is substituted by $e^{-100}$, which was found to be enough to suppress these and to ensure an ensemble of sufficient configurations with all positive eigenvalues. It may be noted that this procedure   effectively introduces a modification of the action, the impact of which is currently not well controlled and requires further investigation in the future.

One of the most important quantities to be calculated from the simulations is the holonomy potential or free energy density $F/V$ at a given temperature. Due to the way it is defined, the calculation through Monte Carlo is not straightforward. However, there is a common method to evaluate it \cite{Larsen:2014yya} which we will adopt here. Note that the only  term that needs to be evaluated from  the Monte Carlo configurations is $e^{-V\mathcal{F}}$ since it is the only one depending upon spatial positions of the dyons/anti-dyons, while all other terms do not have such dependence. In the calculation, according to the standard thermodynamic integration, one introduces an auxiliary parameter $\lambda$ as

\begin{equation}
e^{-V\mathcal{F}_\lambda(\lambda)} = \frac{\int \mathcal{D}r\, e^{-\lambda S_r}}{V^{2(N_L+N_M)}},
\label{Eq:Z_Lambda}
\end{equation} 

where

\begin{equation}
S_r \equiv V_{D\bar{D}} - \log \lsb \det(G_D)\det(G_{\bar{D}})\rsb,
\end{equation}

and $\mathcal{D}r$ is just the integration measure over all dyons' and anti-dyons' positions (for a total of $2(N_L+ N_M)$ of these particles in the simulation). It should be emphasized that for $\lambda=1$, the above \cref{Eq:Z_Lambda} is exactly equal to \cref{Eq:Part_Func3}. The normalization factor $V^{2(N_L+N_M)}$  in the denominator above, is {\em not} introduced arbitrarily but rather comes directly from the construction of the partition function by correctly counting the ``1/V'' factors in the \cref{Eq:Part_Func3}. This proper normalization factor also automatically gives  $\mathcal{F}_\lambda(0)=0$. Then, via standard Monte Carlo simulation procedure, one can compute the ensemble average of the following quantity:  

\begin{equation}
\left\langle S_r \right\rangle(\lambda) \equiv \frac{\int \mathcal{D}r\, S_r\,e^{-\lambda S_r}}{\int \mathcal{D}r\, e^{-\lambda S_r}} = V\frac{\pt \mathcal{F}_\lambda}{\partial\lambda} \,\, .
\end{equation}

Lastly, by integrating out the $\lambda$ dependence of the above, one arrives at the desired free energy:

\begin{equation}
\mathcal{F} = \frac{1}{V}\int_0^1 \text{d}\lambda\left\langle S_r\right\rangle(\lambda) = \mathcal{F}_\lambda(1) ,
\end{equation} 
given that $\mathcal{F}_\lambda(0)=0$ by definition. We emphasize again that for $\lambda=1$, \cref{Eq:Z_Lambda} reduces to \cref{Eq:Part_Func3}, where the denominator $V^{2(N_L+N_M)}$ appears naturally from the construction of the partition function (\cref{Eq:Part_Func,Eq:Part_Func2,Eq:Part_Func3}), allowing to set $\mathcal{F}_\lambda(0)=0$ unambiguously, regardless of the dyon numbers $N_L$ and $N_M$.

Finally, we discuss the choice of the parameters in this framework. For most of the results to be presented, we use a (dimensionless) spatial volume of the box to be $V=43.37$. After several tests, it was determined that the optimal range of (anti)dyon density of each kind was $n_D\in [0,0.5]$ (corresponding to $N_D\in[0,22]$ number of (anti)dyons). Configurations with larger $n_D$ were found to have a rather small contribution to the partition function; therefore, discarded in the simulations (--- see \cref{Sec:Vol_Eff}). We choose the three key parameters as Debye mass $M_D=2$, the core potential strength $V_c=20$ and size $\zeta_j^c=2$; however, in \cref{Sec:Vol_Eff,Sec:Core_Eff,Sec:MD_Eff}, we will vary these quantities to explore the finite volume effects as well as the influence of the three key parameters.

\vspace{0.5cm}

\section{Confinement-Deconfinement Transition}

\subsection{The Holonomy Potential} \label{Sec:FreeEn_Res}

\begin{figure*}
\centering
\begin{minipage}[b]{.49\textwidth}
\includegraphics[width=\linewidth, keepaspectratio]{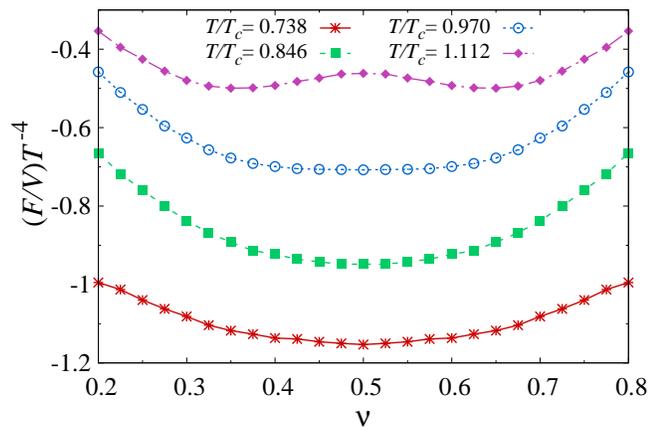}
\end{minipage}\hfill
\begin{minipage}[b]{.49\textwidth}
\includegraphics[width=\linewidth, keepaspectratio]{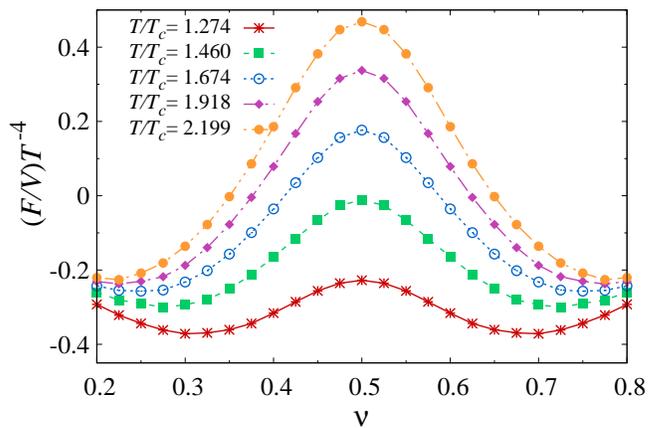}
\end{minipage}
\caption[]{Free energy density $F/V$ as a function of holonomy $\nu$ at various values of temperature. The corresponding action parameters, from bottom top, are $S=5,6,7,8$ (left) and $S=9,10,11,12,13$ (right). }
\label{Fig:FreeEner_S5-13_R3_PBC}
\end{figure*}

It is known from lattice simulations that the $SU(2)$ pure gauge theory has a  certain critical temperature $T_c$, with a confined phase at $T<T_c$, a deconfined phase at $T>T_c$, and a 2nd order phase transition connecting the two phases at $T=T_c$. The relevant order parameter is the expectation value of the Polyakov loop at spatial infinity $L_\infty$ (--- see \cref{AP:HolPol}) which is related to holonomy parameter $\nu$ by $\left\langle L_\infty\right\rangle=\cos(\pi\nu)$. An expectation value of $L_\infty=0$ or $\nu=1/2$ would correspond to the low temperature $Z_2$ center-symmetric, confined phase.   

A first important check is to examine whether such expected phase transition indeed occurs in the dyon ensemble. In order to see that, one needs to compute the holonomy potential, that is, the free energy density as a function of the holonomy $F(\nu)=-T\log\mathcal{Z}$ at varied temperature. Such holonomy potential determines the Polyakov loop dynamics and is a crucial input for a class of chiral models to incorporate confinement dynamics~\cite{Dumitru:2012fw,Lin:2013qu,Smith:2013msa,Pisarski:2016ixt}.  For any given temperature, the minimum of the holonomy potential determines the thermodynamically realized expectation value of the holonomy value which as order parameter thus tells us about the different phases of the theory. As mentioned earlier, the temperature dependence of all the observables in the ensemble comes from the  instanton action $S$ (\cref{Eq:S_Inst}) which is an input parameter in the simulation. \cref{Fig:FreeEner_S5-13_R3_PBC} shows the free energy density for $S=5,6,\ldots,13$. It is found that for $S=5\sim7$, the minimum of the free energy density lies at $\nu_{\text{min}}=0.5$, namely maximal non-trivial holonomy corresponding to the confined phase. For $S>7$, the shape of $F/V$ becomes that of a symmetric double well potential with two minima located at $\nu_{\text{min}}<0.5$ and $\nu_{\text{min}}>0.5$ in a symmetric way. It shall be mentioned that as expected for an $SU(2)$ pure gauge theory, $F/V$ is symmetric under the interchange $\nu\to\bar{\nu}=1-\nu$, and this feature has indeed been validated explicitly in the numerical calculations. So the results clearly reveal a confined phase at low temperature while a deconfined phase at high temperature. 

\begin{figure}[t]
\centering
\includegraphics[width=\textwidth]{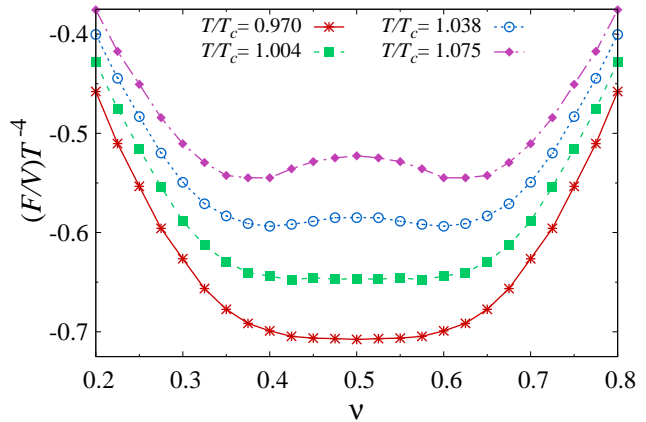}
\caption{Free energy density $F/V$ as a function of holonomy $\nu$ at several values of temperature near the phase transition point, with the corresponding action parameters $S=7.00, 7.25, 7.50, 7.75$, from bottom to top.}
\label{Fig:FreeEner_Tc_R3_PBC}
\end{figure}

To more accurately locate the critical action (or equivalently the critical temperature $T_c$), we further run the simulation for $S=7.25, 7.5$ and $7.75$. As shown on \cref{Fig:FreeEner_Tc_R3_PBC}, for $S\geq7.5$ the $Z_2$ symmetry is clearly broken and the minimum of the free energy density is shifted away from the $\nu=\bar{\nu}=0.5$. For $S=7.25$, more points were necessary to examine the minimum, and despite the potential on \cref{Fig:FreeEner_Tc_R3_PBC} exhibits a very flat dependence around $\nu=0.5$, the minimum was actually found around $\nu_{\text{min}}\approx0.453$. Thus, at the present numerical precision, we determine the critical temperature at $S_c=7.22$, which fixes our scale parameter from \cref{Eq:S_Inst} at $\Lambda=0.373T_c$ and allows us to express all temperature dependent quantities in terms of the ratio $T/T_c$. 

We next come to the expectation value $\left\langle L_\infty\right\rangle = \cos(\pi \nu)$, which can be determined from the position of the minimum of the holonomy potential. This is done by fitting the free energy density near the minima to a quadratic function with 9 to 15 points and then, through a derivative test on the fit, finding its minimum accurately. 

As an important and insightful check of the role of $L_\infty$ as an order parameter for the expected 2nd order phase transition, we quantitatively examine whether its dependence on temperature near the transition point  follows the proper  universality class. The well known Svetitsky-Yaffe conjecture \cite{Svetitsky:1982gs}, relates $SU(2)$ pure gauge theory in $(3+1)$ dimensions to the $3D$ Ising model of ferromagnetism by categorizing both in the same universality class, which has been proven several times in different numerical studies such as \cite{Fiore:1998uk,Engels:1989fz,Engels:1995em,Engels:1992fs,Teper:1993gp,Caselle:1995wn}. In this sense, $L_\infty$ becomes the analog of the magnetization, thus its critical behavior must follow the same universal power law

\begin{equation}
\left\langle L_\infty\right\rangle = b(T/T_c -1)^\beta \lsb 1+d(T/T_c-1)^\Delta\rsb,
\label{Eq:LCritical}
\end{equation}
with $b$ and $d$ the fitting parameters.

\begin{figure}[]
\centering
\includegraphics[width=1\textwidth]{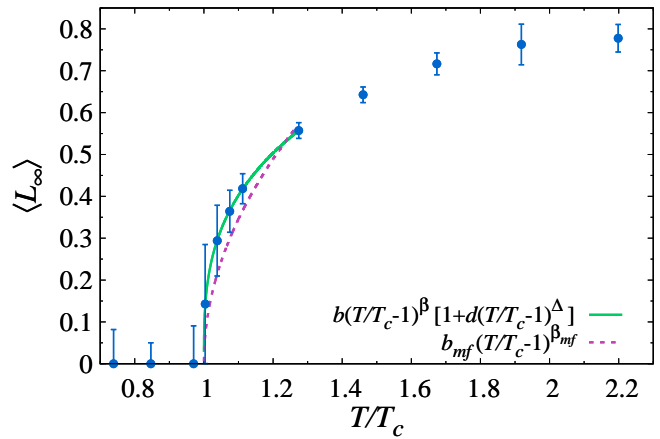}
\caption{$L_\infty$ as an order parameter of the 2nd order phase transition with $b=0.858(142)$ and $d=-0.017(348)$ as the fitting parameters to the power law \cref{Eq:LCritical} and $b_{mf}=1.094(33)$ the corresponding one to the mean-field fit.}
\label{Fig:L-OP}
\end{figure}

Using the well established values of the critical exponents of the $3D$ Ising model $\beta\approx 0.3265(3)$ and $\Delta\approx0.530(16)$ \cite{Pelissetto:2000ek}, on \cref{Fig:L-OP} we show the fitted curve obtained from the numerical results of the dyon ensemble in the near-$T_c$ region, namely $1\leq T/T_c\leq 1.274$. The very low value of $\chi^2=1.44\times10^{-4}$ of the fit (which is partly due to the sizable error bar because of limited statistics) suggests an almost perfect agreement between the confinement/deconifnemnt phase transition behavior with the anticipated critical behavior of the $3D$ Ising model's 2nd order phase transition. It also demonstrates qualitative agreement with the lattice results from \cite{Digal:2003jc,Hubner:2008ef}. For completeness and comparison, we also show the fit using the mean-field critical exponent $\beta_{mf}=1/2$, which shows a qualitatively similar trend but a significantly  larger value of $\chi^2=0.13$. The comparison favors the former fitting and implies that the transition from the dyon ensembles captures the beyond-mean-field critical behavior of a 2nd order phase transition.

\begin{figure}[]
\centering
\includegraphics[width=1\textwidth]{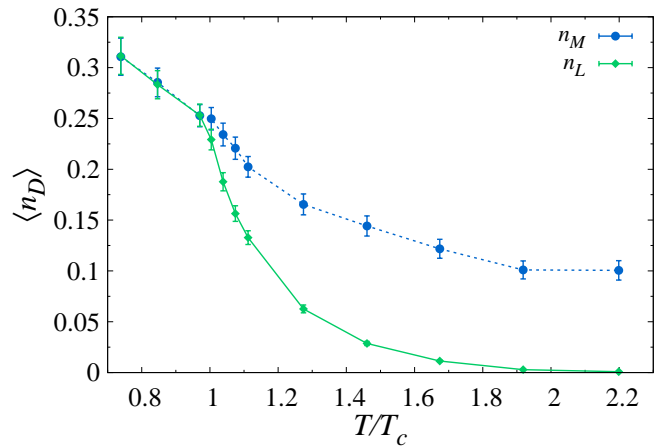}
\caption{Temperature dependence of the ensemble average of dyon densities.}
\label{Fig:D_Densities}
\end{figure}

We now report the results for the expectation values of dyon densities, shown in \cref{Fig:D_Densities}.  One can see that at $T< T_c$, the $L$ and $M$ type densities are equal as expected. In the confined phase, the preferred holonomy corresponds to the maximally non-trivial one where both dyon types have the same core radius as well as equal action share and therefore equal weight in the partition function. For $T>T_c$, the prefered holonomy starts to shift away from the symmetric point towards the trivial holonomy ($\nu\to 0$ in this case) and the $M$ dyons become larger and larger. Recalling from the KvBLL caloron solution (--- see \cref{AP:KvBLL}), in the limit of trivial holonomy, the $L$ dyon disappears and the field becomes that of the Harrington-Shepard caloron. A similar situation is observed in the ensemble as temperature is increased, with the $L$ dyon density decreasing much faster than $M$ type. The total density of all these magnetically charged objects demonstrates a strong temperature dependence with very rapid increase from high temperature toward near $T_c$ regime, in consistency with the magnetic scenario.

\subsection{The Dyon and Anti-dyon Spatial Correlations}

\begin{figure}[]
\centering
\includegraphics[width=1\textwidth]{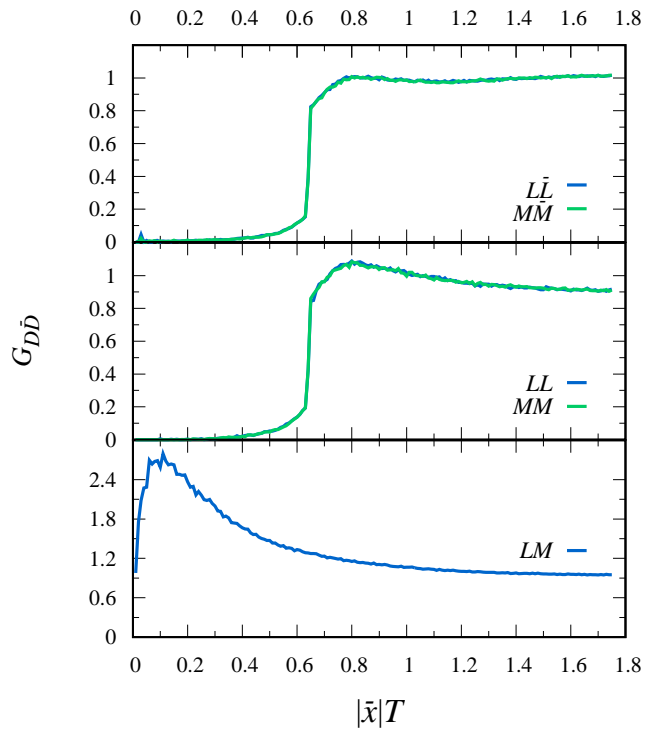}
\caption{2-particle spatial correlations for dyon--dyon and dyon--anti-dyon pairs at $T<T_c$, namely, in the confined phase.}
\label{Fig:GDD-S5}
\end{figure}

\begin{figure}[]
\centering
\includegraphics[width=1\textwidth]{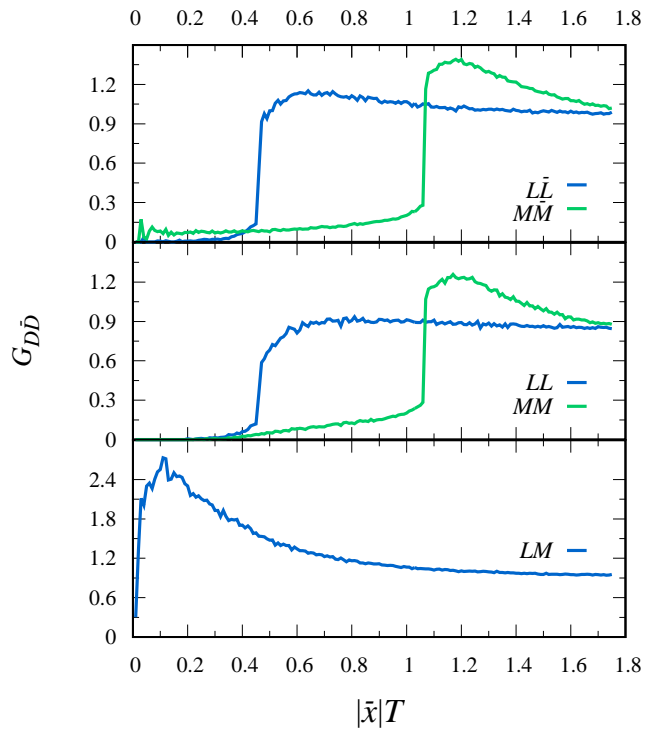}
\caption{2-particle spatial correlations for dyon--dyon and dyon--anti-dyon pairs at $T>T_c$, namely, in the deconfined phase.}
\label{Fig:GDD-S9}
\end{figure}

The interactions between dyons and anti-dyons are essential for the properties of the ensemble and in particular for driving the system toward confinement at high dyon density (i.e. low temperature) regime. The effect of such interactions can be illustrated by examining the spatial density density correlations among various pairs of dyons/anti-dyons, as defined in the following: 

\begin{equation}
G_{DD'}(|\vec{x}|)=\frac{\frac{1}{V}\left\lan \sum\limits_{i=1}^{N_D}\sum\limits_{j=1}^{N_{D'}}\Theta_{\delta x}(r_{ij}-|\vec{x}|) \right\ran}{n_D n_{D'}\frac{4\pi}{3}\lsb(|\vec{x}|+\delta x)^3-|\vec{x}|^3\rsb},
\end{equation}
which is normalized to that of an uncorrelated ideal gas and where the step function $\Theta_{\delta x}(\xi)=1$ for $0<\xi<\delta x$ and 0 otherwise. A value of unity for the $G_{DD'}$ would indicate a situation without correlations as is characteristic for a free gas ensemble. The numerical results are shown in \cref{Fig:GDD-S5,Fig:GDD-S9} for all different (anti)dyon pairs combinations, each computed at one temperature value below $T_c$ and another one above $T_c$. These results are obtained under equilibrium conditions for given temperature, i.e. with  holonomy parameter $\nu$ being the one at minimal free energy and the number of dyons $N_D$ fixed at  the ensemble averaged values. 

The presence of the repulsive core is clearly observed for all the dyon pairs, besides the $LM$ for which there is none. At distances right above the core size $\zeta_j^c/2\pi\nu_j$ the correlation functions seem to have a small bump that rapidly goes to unity at larger distance, indicating at a short-range correlation pattern arising from the repulsive core. 

\subsection{The Polyakov Loop Correlator}

Besides the Polyakov loop expectation value itself, another important ``indicator'' of the confinement/deconfinement transition is the static (quark-anti-quark) potential which essentially is evaluated from a temporal Wilson loop or equivalently the spatial correlator of the Polyakov loop. In particular the so-obtained static potential is expected, at large spatial separation, to exhibit a linearly rising behavior  in the confined phase while to level off in the deconfined phase. It is important to evaluate this observable in the dyon ensemble. 

The computation is however technically tricky in the present framework. In the large distance limit ($|\vec{x}|\to\infty$), the $A_4$ component of the dyon fields becomes Abelian (--- see \cref{AP:KvBLL}). However, the total $A_4$ of the ensemble (far away from their individual cores) cannot be given by a superposition of the individual fields of all dyons yet. Since the asymptotic condition $A_4|_{|\vec{x}|\to\infty}=\pi\nu\tau^3$ must be satisfied, one has to eliminate the holonomy parameter term in the gauge field associated with individual dyon by means of the time dependent gauge transformation $U=\exp(-i\pi\nu x_4\tau^3)$, after which one can then superimpose all dyonic fields and finally restore the asymptotic term with the inverse gauge transformation $U^\dagger$ \cite{Larsen:2014yya}. This procedure leads to

\begin{equation}
A_4(\vec{x})=\frac{\tau^3}{2}\lsb 2\pi\nu + l(\vec{x})\rsb,  
\label{Eq:A4_Total}
\end{equation} 
where $l(\vec{x})$ is the sum of all Coulomb terms of dyons and anti-dyons

\begin{flalign}
\nb
l(\vec{x}) &\equiv  \sum_{l,m}^{N_L,N_M}\lb \frac{1}{|\vec{x}-\vec{r}_{L_l}|} - \frac{1}{|\vec{x}-\vec{r}_{M_m}|}\right.\\
&+ \left.\frac{1}{|\vec{x}-\vec{r}_{\bar{L}_l}|} - \frac{1}{|\vec{x}-\vec{r}_{\bar{M}_m}|} \rb.
\label{Eq:A4_l(x)}
\end{flalign}

At finite temperature, the color averaged heavy quark-antiquark free energy $F^{\text{avg}}_{q\bar{q}}$ is defined through the expectation value of traced Polyakov loop correlators (--- see \cref{AP:HolPol}). For quarks in the fundamental representation, from \cref{Eq:A4_Total,Eq:A4_l(x)} and the definition of the Polyakov loop, it is straightforward to see that

\begin{equation}
\frac{1}{2}\Tr \,L^f(\vec{x})=\cos\lsb \pi\nu + \frac{1}{2}l(\vec{x})\rsb \, .
\end{equation}
Thus, the color averaged static quark-anti-quark potential in the dyon ensemble is given by 

\begin{flalign}
\nb
e^{-F^{\text{avg}}_{q\bar{q}}}&\equiv   \frac{1}{4}\left\lan\Tr\, L^{\dagger f}(\vec{x})\Tr \,L^f(\vec{y})\right\ran \\
\nb
&= \left\lan\cos\lsb \pi\nu + \frac{1}{2}l(\vec{x})\rsb \cos\lsb \pi\nu + \frac{1}{2}l(\vec{y})\rsb\right\ran.\\
\end{flalign} 
The above static potential, though, is different from a color-singlet static potential which is the one relevant for linear behavior at large separation. According to the color decomposition $2\otimes\bar{2}=1\oplus3$, an $SU(2)$ quark-antiquark pair can interact through a singlet and a triplet channel \cite{Nadkarni:1986as}, meaning that $F^{\text{avg}}_{q\bar{q}}$ is decomposed into

\begin{equation}
e^{F_{q\bar{q}}^{\text{avg}}}=\frac{1}{4}e^{-F^1_{q\bar{q}}} + \frac{3}{4}e^{-F^3_{q\bar{q}}}, 
\label{Eq:Fqqbar_Decomp}
\end{equation} 
where the singlet free energy is obtained from the following:

\begin{flalign}
\nb
e^{-F^1_{q\bar{q}}}&\equiv  \frac{1}{2}\left\lan\Tr\lsb L^{\dagger f}(\vec{x})L^f(\vec{y})\rsb\right\ran \\
&= \left\lan\cos\lsb \frac{l(\vec{x})-l(\vec{y})}{2}\rsb\right\ran
\end{flalign} 
and the triplet contribution follows trivially from \cref{Eq:Fqqbar_Decomp}. 

\begin{figure}[]
\centering
\includegraphics[width=1\textwidth]{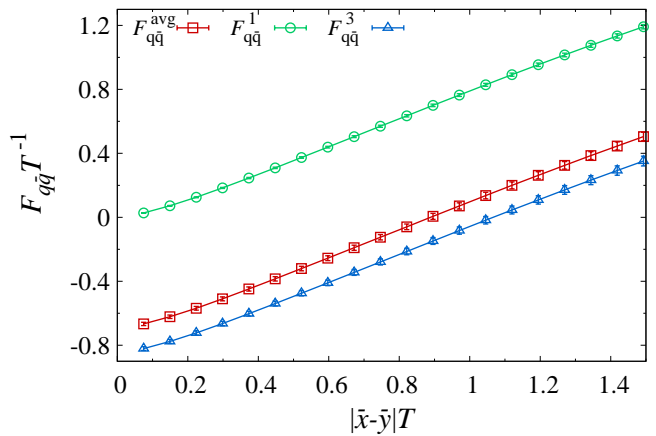}
\caption{The color-averaged  static quark-anti-quark potential as well as its decomposed singlet and triplet channel components in the confined phase at $T/T_c=0.970$ and $\nu=0.5$.}
\label{Fig:FqqCAve1}
\end{figure}

\begin{figure}[]
\centering
\includegraphics[width=1\textwidth]{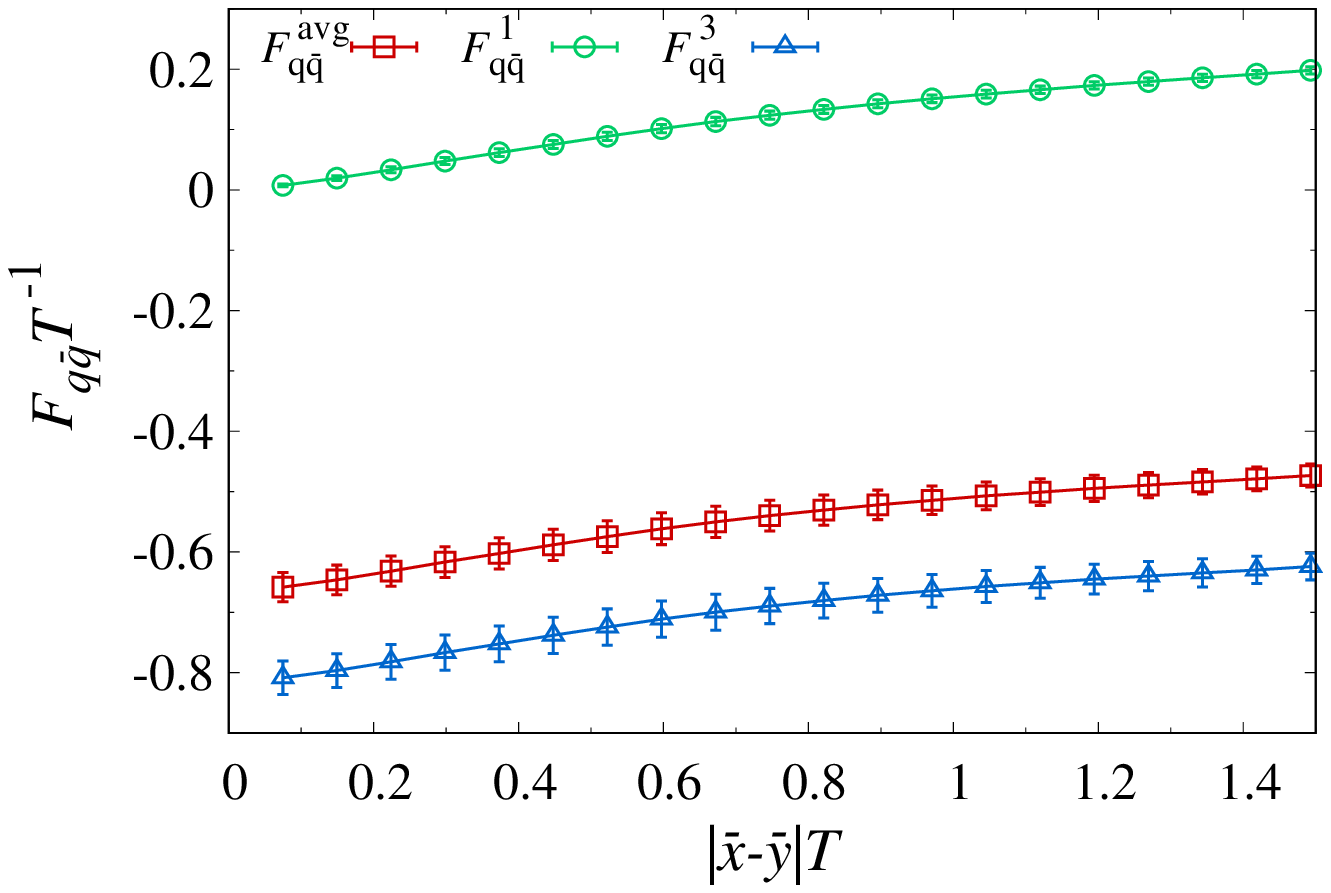}
\caption{The color-averaged  static quark-anti-quark potential as well as its decomposed singlet and triplet channel components  in the deconfined phase at $T/T_c=1.674$ and $\nu=0.250$.}
\label{Fig:FqqCAve2}
\end{figure}

\begin{figure}[]
\centering
\includegraphics[width=1\textwidth]{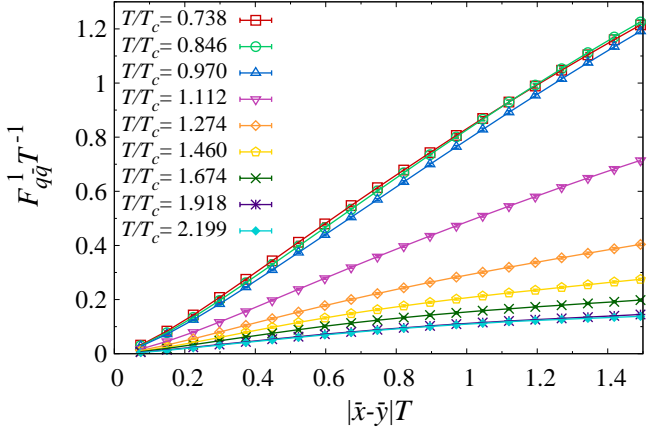}
\caption{The singlet channel of the static quark-anti-quark potential for the fundamental representation in both confined and deconfined phases for $\nu=0.5, 0.5, 0.5, 0.35, 0.3, 0.275, 0.25, 0.225$ and 0.225, in order of increasing temperature.}
\label{Fig:FqqSing}
\end{figure}

Due to the periodic boundary conditions imposed in our geometry, the maximum allowed distance is $|\vec{x}-\vec{y}|\leq R/2$, where $R\approx 3.51$ is the size of the box of volume $V=43.37$. To compute these observables, a total of 3000 Monte Carlo configurations are used for each combination of number of dyons ($N_L,N_M=0,...,22$). To account for isotropy, for each interquark separation we averaged the contribution to the Polyakov loop correlator from 13 different orientations. 
At each temperature, the  holonomy parameter $\nu$ is fixed to be the equilibrium value that the one which minimizes  the ensemble free energy.  In \cref{Fig:FqqCAve1,Fig:FqqCAve2}, we show the color averaged potential and its singlet and triplet contributions for the confined and deconfined phases at $T/T_c=0.970$ and $T/T_c=1.674$ respectively.  In \cref{Fig:FqqSing} we show the singlet channel free energy alone as a function of interquark separation $|\vec{x}-\vec{y}|$ for several temperatures below and above $T_c$. It may be noted that the color-averaged static potential above $T_c$ appears not fully saturated at large distance, due to two factors. The first is the finite volume effect (as will be discussed later in Section IV.A) which would limit the largest possible distance we could explore.  The second is that in the high temperature deconfined phase, the perturbative thermal gluons (which are absent in the current framework)  would contribute more and more importantly with increasing temperature to the screening of the static potentials.

\begin{figure}[]
\centering
\includegraphics[width=1\textwidth]{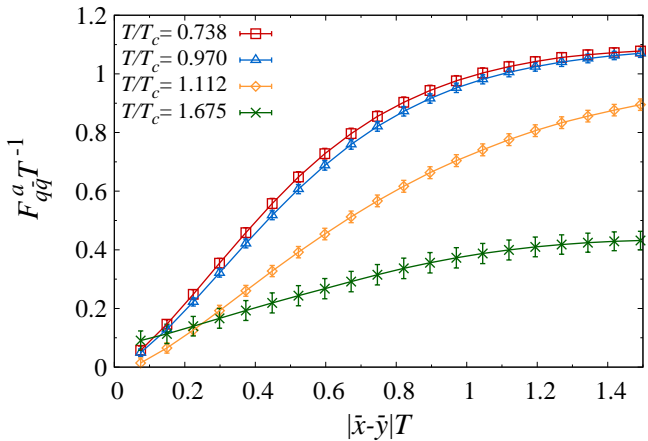}
\caption{The static quark-anti-quark potential for the adjoint representation in both confined and deconfined phases for $\nu=0.5, 0.5, 0.3$ and 0.25, in order of increasing temperature.}
\label{Fig:FqqAdj}
\end{figure}

An interesting comparison is with the static potential of quarks and anti-quarks in the adjoint representation, in which case no linear rising at large distance is expected as the gluons (themselves being adjoint) can screen out the potential.   Following \cite{Smith:2013msa}, one can obtain the adjoint static potential via the following relation with  the fundamental one $L^f$: 

\begin{equation}
L_{ij}^a=\frac{1}{2}\Tr\lb\tau_i L^f\tau_j L^{\dagger f} \rb, 
\label{Eq:Pol_Adj1}
\end{equation}
where $\tau_i$ are the Pauli matrices. Given that $L\in SU(2)$, the fundamental representation is generally defined as

\begin{equation}
L^f = a_0 \mathds{1} + ia_j\tau_j, 
\end{equation} 
with $a_\mu a_\mu=1$. Thus \cref{Eq:Pol_Adj1} can be rewritten as

\begin{equation}
L^a_{ij}=2 \lsb a_0a_k\varepsilon_{ijk} + a_ia_j + \delta_{ij}\lb a_0^2 -\frac{1}{2}\rb \rsb, 
\label{Eq:Pol_Adj2}
\end{equation}
and it is easy to see that its trace is expressed in terms of the fundamental one as

\begin{equation}
\Tr \,L^a(\vec{x}) = \left|\Tr \,L^f(\vec{x})\right|^2-1. 
\end{equation}
Therefore the adjoint static quark-antiquark free energy is then given by

\begin{equation}
e^{-F_{q\bar{q}}^{a}}=\frac{\left\lan\Tr \,L^{\dagger a}(\vec{x})\Tr \,L^a(\vec{y})\right\ran}{\left\lan \left|\Tr \,L^a(0)\right|^2 \right\ran}. 
\end{equation}
Notice we have included a normalization factor $\left\lan \left|\Tr \,L^a(0)\right|^2 \right\ran$ in the correlator such that $F_{q\bar{q}}^a=0$ at $|\vec{x}-\vec{y}|=0$. The resulting potentials are shown in \cref{Fig:FqqAdj} for different temperatures.

\begin{figure}[]
\centering
\includegraphics[width=\textwidth]{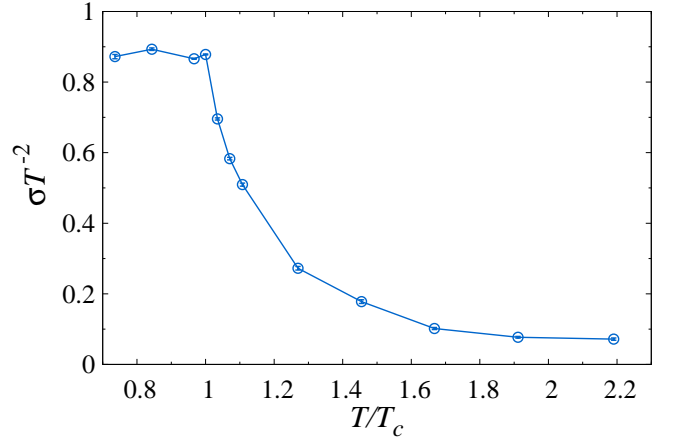}
\caption{Temperature dependence of the string tension $\sigma$ extracted from the singlet static quark-anti-quark potential $F^1_{q\bar{q}}$ in the fundamental representation.}
\label{Fig:StrTsn_Avg}
\end{figure}

As pointed out already, at large separation in the confined phase, one expects the (fundamental representation) singlet static potential to have a linear rising behavior of the following form: 

\begin{equation}
\left. F_{q\bar{q}}\right|_{|\vec{x}-\vec{y}|\to\infty} \approx \sigma |\vec{x}-\vec{y}|,  
\label{Eq:Fqqbar}
\end{equation}
where $\sigma$ is the so called \textit{string tension}. This is clearly observed in the fundamental representation (\cref{Fig:FqqSing}) at $T/T_c < 1$. However, at temperatures above $T_c$, the slope $\sigma$ drops toward zero, as expected. In \cref{Fig:StrTsn_Avg}, we show the extracted string tensions for the singlet potential for several temperatures. In extracting the slope, we use linear fits for the large distance part but  ignore the ``curved" tails observed at the largest distances, which are most likely  due to finite volume effects (--- see \cref{Sec:Vol_Eff} for an extended discussion). It may also be mentioned that our results show a relatively slow decrease of $\sigma$ above $T_c$, an effect which may also be due to finite volume issues. In contrast, the  adjoint static potential does not show any linear rising at large separation.  In short, our results from the dyon ensemble for the static quark-antiquark potentials in both representations are consistent with the expected behavior of an $SU(2)$ pure gauge theory.

\subsection{The Spatial Wilson Loop}

Another interesting quantity to explore is the spatial Wilson loop. It is known that the spatial Wilson loop at finite temperature shows area law behavior with a finite spatial string tension $\sigma_s$ both below and above $T_c$ and thus by itself does not serve as an ``indicator'' of confinement transition~\cite{Borgs:1985qh,Manousakis:1986jh,Bali:1993tz,Karkkainen:1993ch}. Nonetheless, the restoration of Lorentz symmetry (Euclidean $O(4)$) at $T\to0$ suggests that in this limit, $\sigma_s$ should coincide with the string tension of the static potential \cref{Eq:Fqqbar} extracted from Polyakov loop correlators. The $SU(2)$ traced spatial Wilson loop is defined as

\begin{equation}
W_C \equiv \frac{1}{2}\Tr\,\mathcal{P}\exp  \lsb i\oint_C \diff x_i A_i(x) \rsb. 
\end{equation} 

In the gauge where $A_4(x)$ is diagonal (--- see \cref{AP:KvBLL}), the only non-vanishing spatial component of the dyon fields, in the asymptotic limit, is 

\begin{equation}
A_\phi^j (\vec{x})= m_j\frac{\tan \frac{\theta}{2}}{r}\frac{\tau^3}{2}, 
\label{Eq:Aphi}
\end{equation}
where $m_j=\pm1$ is the corresponding magnetic charge (\cref{Tab:Dyon_Prop}) and $r=\sqrt{x_ix_i}$. The Dirac string singularity along the negative $x_3$-axis, although a gauge artifact, might be an inconvenience for the numerical simulations. Therefore, for computing $W_C$ it is more suitable to use the corresponding magnetic field  (instead of the gauge potential). For this, the Abelian Stokes theorem can be used to rewrite the spatial Wilson Loop in the so called ``Abelian dominance" approximation \cite{Greensite:2014gra} 

\begin{equation}
W_C \approx  \frac{1}{2}\Tr\,\exp  \lsb i\int_{A_C} \diff a_i B_i(x) \rsb, 
\end{equation}
where $B_i\equiv\frac{1}{2}\varepsilon_{ijk}F_{jk}$ and $A_C$ is the area enclosed by a rectangular contour $C$. Therefore, the corresponding magnetic field to \cref{Eq:Aphi} is

\begin{equation}
B^j_r= \frac{m_j}{r^2}\frac{\tau^3}{2}, 
\end{equation}
and $B_\phi=B_\theta=0$. The total field strength from the whole ensemble will thus be

\begin{flalign}
\nb
B_i(\vec{x}) &= \frac{\tau^3}{2}\sum_{l,m}^{N_L,N_M}\lsb \frac{(\vec{x}-\vec{r}_{L_l})_i}{|\vec{x}-\vec{r}_{L_l}|^3} - \frac{(\vec{x}-\vec{r}_{M_m})_i}{|\vec{x}-\vec{r}_{M_m}|^3} \right.\\
& \left. - \frac{(\vec{x}-\vec{r}_{\bar{L}_l})_i}{|\vec{x}-\vec{r}_{\bar{L}_l}|^3} + \frac{(\vec{x}-\vec{r}_{\bar{M}_m})_i}{|\vec{x}-\vec{r}_{\bar{M}_m}|^3} \rsb.
\end{flalign}

\begin{figure}[]
\centering
\includegraphics[width=1\textwidth]{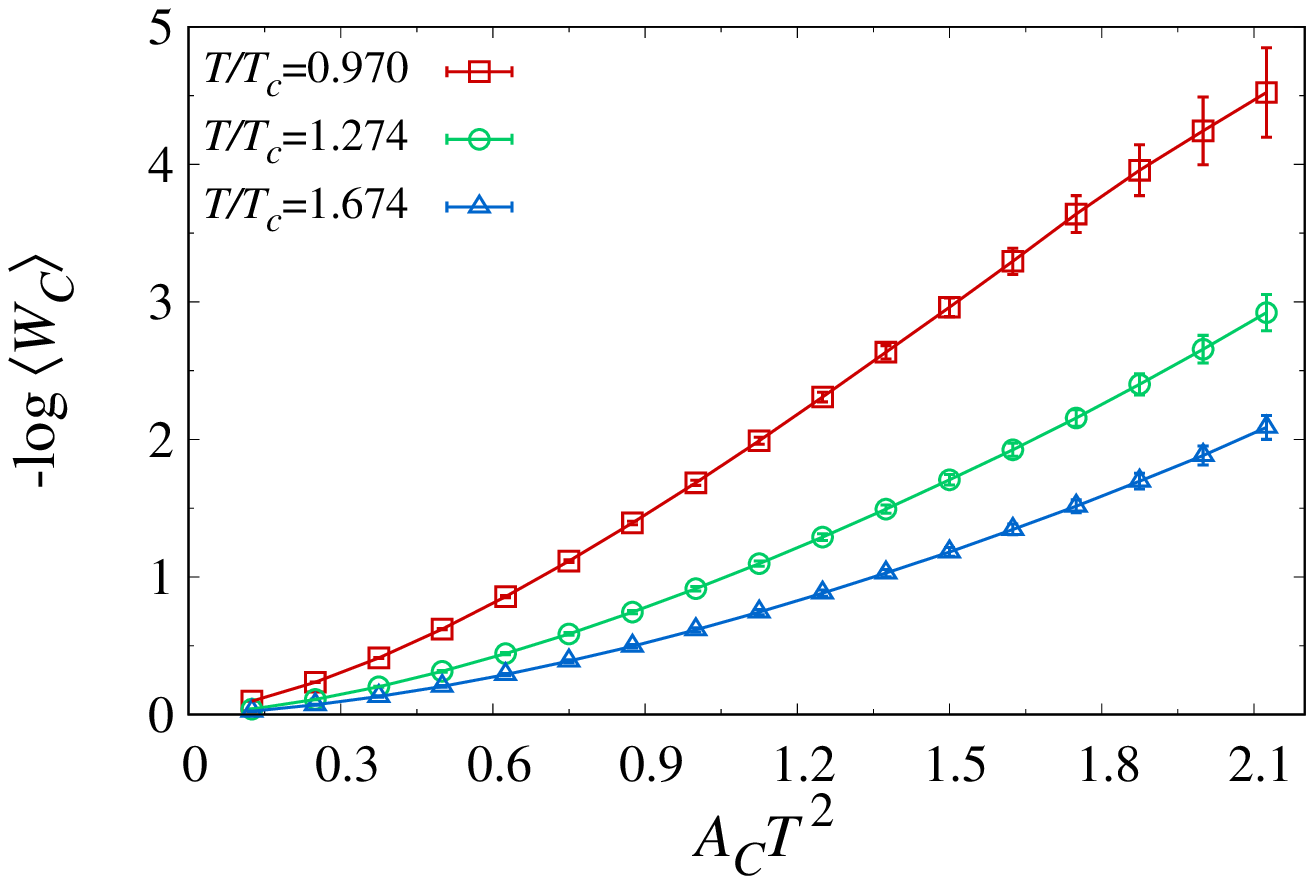}
\caption{The spatial Wilson loop (in the form of its negative logarithm) for the fundamental representation in both confined and deconfined phases.}
\label{Fig:Wilson_Fund}
\end{figure}

\begin{figure}[]
\centering
\includegraphics[width=1\textwidth]{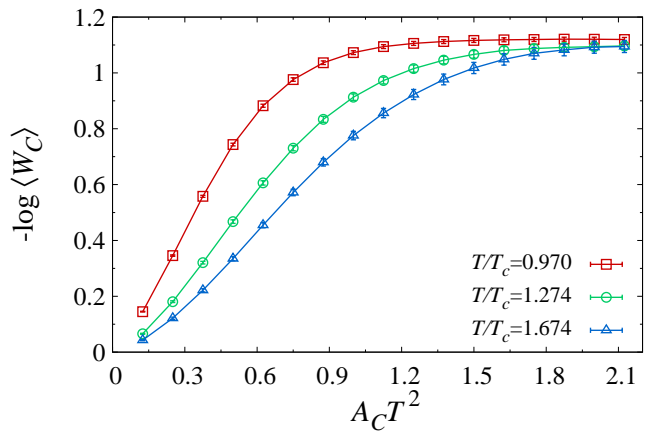}
\caption{The spatial Wilson loop (in the form of its negative logarithm) for the adjoint representation in both confined and deconfined phases.}
\label{Fig:Wilson_Adj}
\end{figure}

It is interesting to examine whether the spatial Wilson loop computed from the dyon ensemble will follow the area law in both confining and deconfined phases, i.e.

\begin{equation}
\left\langle W_C \right\rangle \propto e^{-\sigma_s A_C} \, .
\end{equation} 
 In \cref{Fig:Wilson_Fund}, the negative logarithm of $\left\langle W_C \right\rangle$ in the fundamental representation is plotted as a function of the spatial loop area $A_C$, which indeed demonstrates an almost linear rising behavior at large contour areas.  

Recalling the units used in this work, the string tension obtained here is dimensionless after rescaling all quantities by temperature. To restore physical units, one makes the change  $\sigma_s\to\sigma_s/T^2$. As has been established before~ \cite{Bali:1993tz}, $\sigma_s$ increases with increasing $T$, however, $\sigma_s/T^2$ should decrease with increasing temperature. Such a trend is consistent with  our results from dyon ensemble. Finally, we've also examined the spatial Wilson loop for the adjoint representation, shown in \cref{Fig:Wilson_Adj}. It is observed that the curve  rises rapidly with loop area and reaches a plateau much faster than that of $F_{q\bar{q}}^a$, again an indication of the screening effects for adjoint sources.

\section{Discussions}

\subsection{Finite volume effects} \label{Sec:Vol_Eff}

A rigorous study of all thermodynamic quantities in principle requires an infinite volume limit, which is obviously impossible for any realistic numerical simulations. In the case of the present study on the dyon ensemble, using a larger volume requires an increased number of dyons/anti-dyons in the simulations thus costing significantly more computing power. A practical approach would be to examine the finite size effect by  perform tests with increasing volume of the box. In this Subsection we compare results obtained with  two and three times the originally used volume, denoted as $V_0=43.37$.

\begin{figure*}
\centering
\begin{minipage}[b]{.33\textwidth}
\includegraphics[width=\linewidth, keepaspectratio]{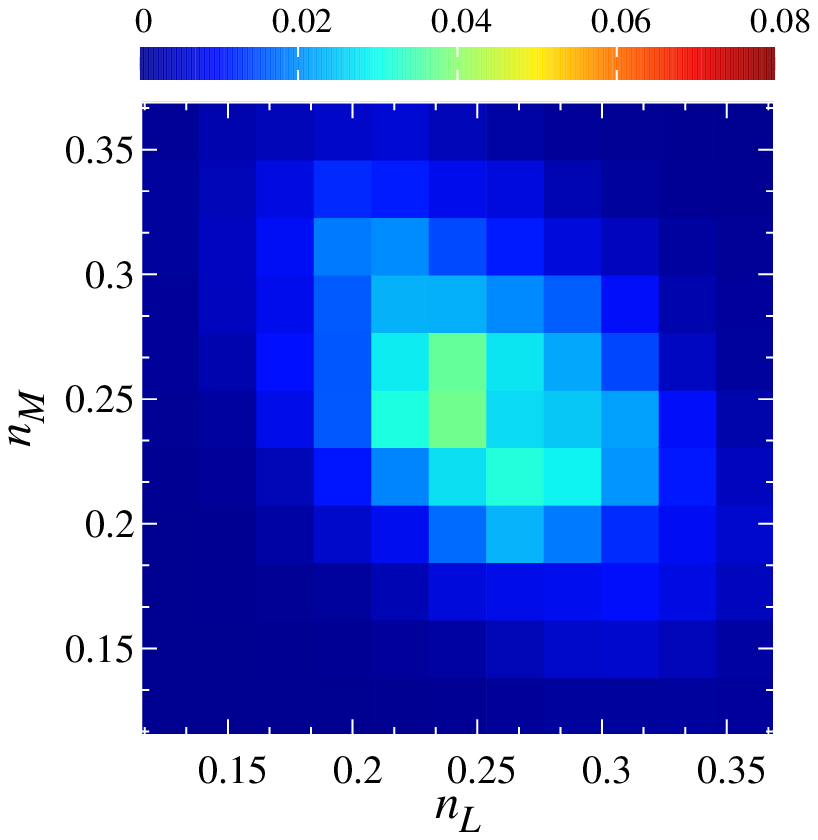}
\end{minipage}\hfill
\begin{minipage}[b]{.33\textwidth}
\includegraphics[width=\linewidth, keepaspectratio]{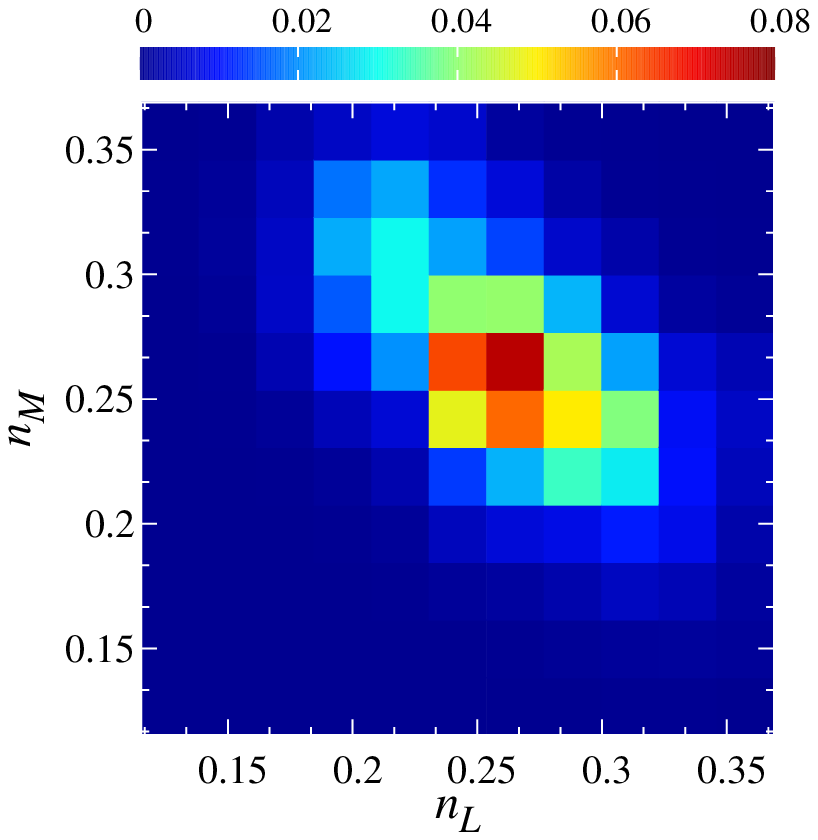}
\end{minipage}\hfill
\begin{minipage}[b]{.33\textwidth}
\includegraphics[width=\linewidth, keepaspectratio]{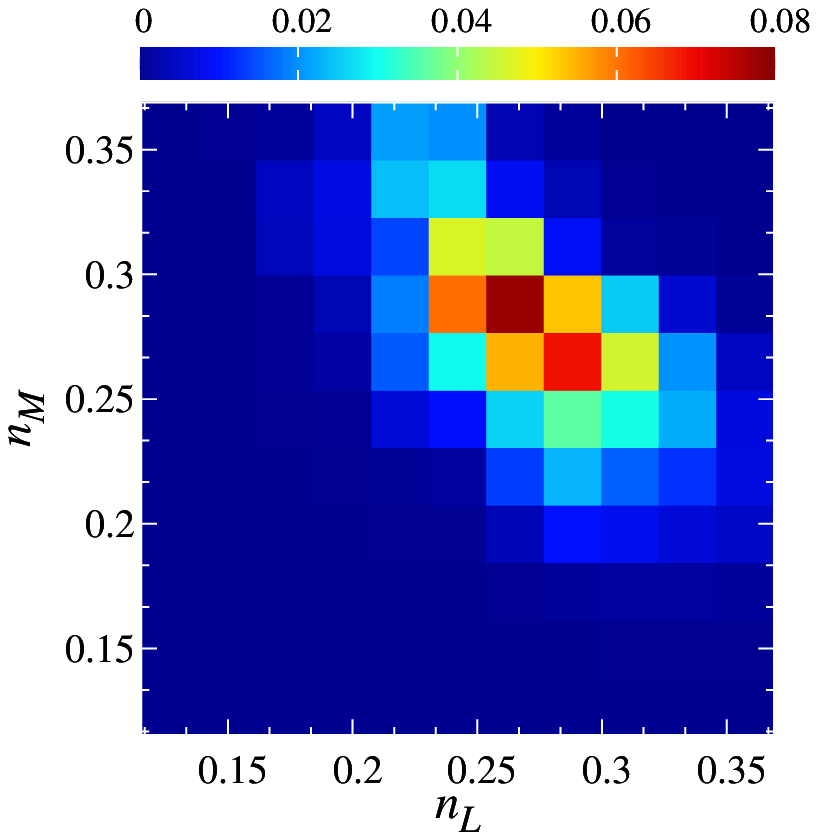}
\end{minipage}
\caption{Dyon density dependence of $\mathcal{Z}_{\text{LM}}/\mathcal{Z}$ for different volumes (from left to right: $V_0, 2V_0 \text{ and } 3V_0$) in the confined phase $T/T_c=0.970$ and $\nu=0.5$.}
\label{Fig:NDistVol_S7_R3_PBC}
\end{figure*}

\begin{figure*}
\centering
\begin{minipage}[b]{.33\textwidth}
\includegraphics[width=\linewidth, keepaspectratio]{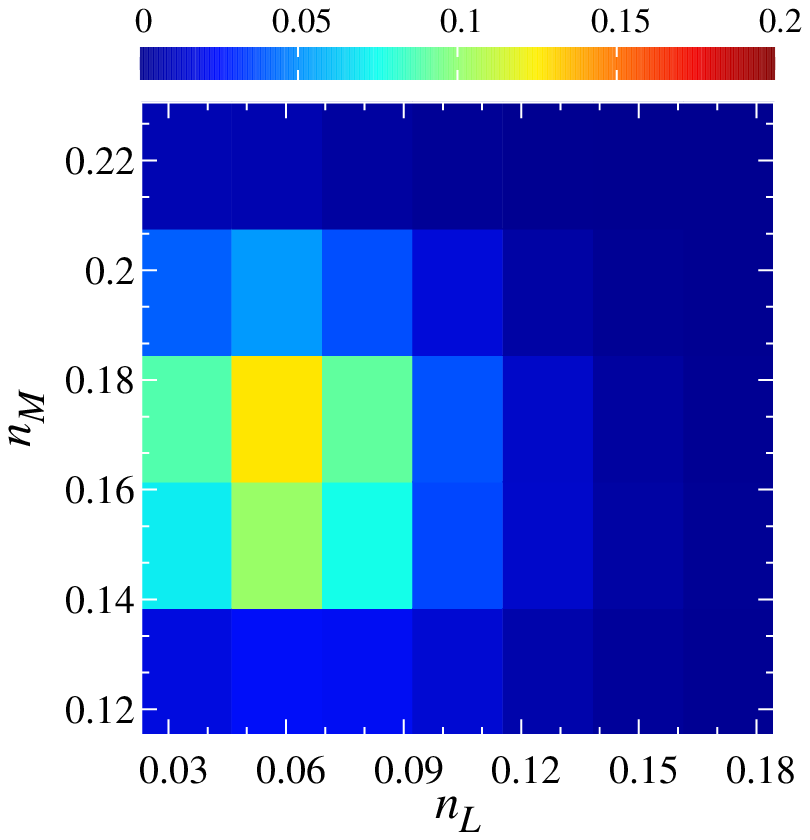}
\end{minipage}\hfill
\begin{minipage}[b]{.33\textwidth}
\includegraphics[width=\linewidth, keepaspectratio]{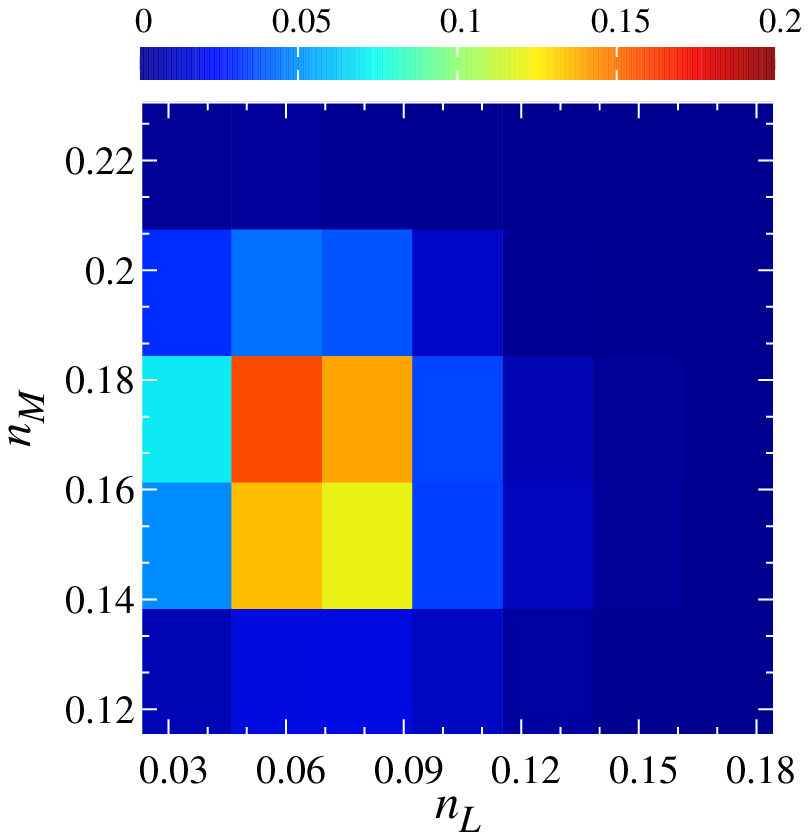}
\end{minipage}\hfill
\begin{minipage}[b]{.33\textwidth}
\includegraphics[width=\linewidth, keepaspectratio]{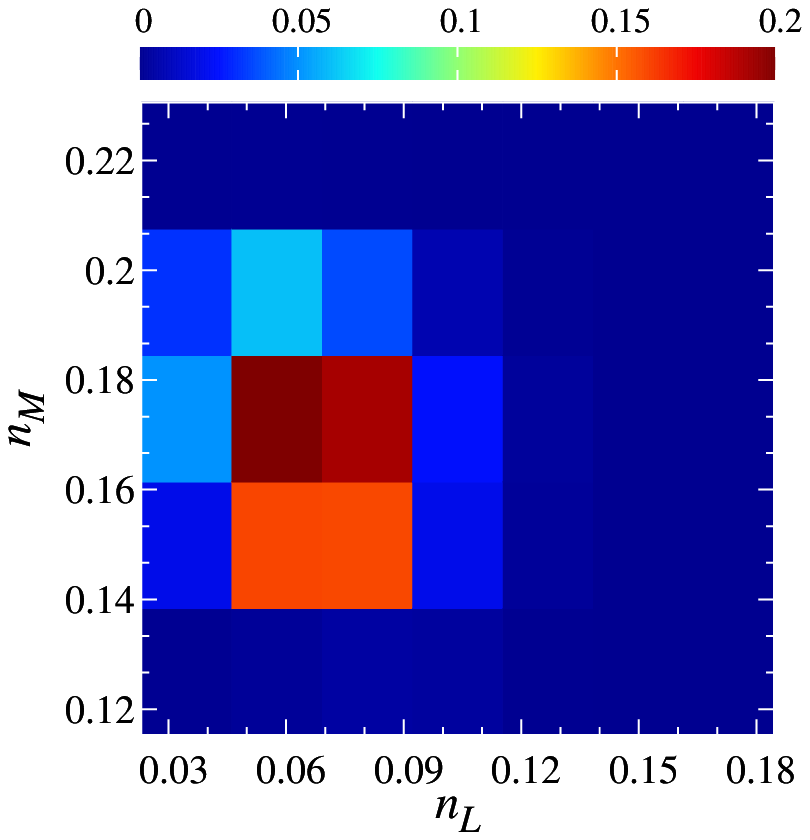}
\end{minipage}
\caption{Dyon density dependence of $\mathcal{Z}_{\text{LM}}/\mathcal{Z}$ for different volumes (from left to right: $V_0, 2V_0 \text{ and } 3V_0$) in the deconfined phase $T/T_c=1.274$ and $\nu=0.3$.}
\label{Fig:NDistVol_S9_R3_PBC}
\end{figure*}

One important feature to check is the (relative) contribution from various terms $\mathcal{Z}_{\text{LM}}$ in the partition function $\mathcal{Z}$ \cref{Eq:Part_Func4}. Note that for different volumes, each term with fixed number of dyons/anti-dyons $N_{L,M}$ would have different density. The better way to compare results computed with different volume would be to examine the contribution from given dyon/anti-dyon densities.  In \cref{Fig:NDistVol_S7_R3_PBC,Fig:NDistVol_S9_R3_PBC}, we show how the contribution to partition function (from individual fixed-density terms in the ensemble sum)  is changed with the increased volume. As expected, the maximum peak of the distribution becomes sharper around the most probable densities.  Most importantly the location of the maximum does not change much with increased volume. \cref{Tab:Vol_Eff} summarizes and compares numerical values of ensemble averages of the dyon densities at different temperatures as well as the free energy density for $\nu_{\text{min}}$  obtained at the three different volumes. It can be seen that going from $V_0$ up to $3V_0$, there is a small shift (at few percent level) of the free energy density while small changes in the dyon densities. Such comparison clearly demonstrates that our thermodynamic results are quite stable with increasing system volume, which is an indication that our results shall be a very good approximation to the thermodynamic limit. 

\begin{table}
\begin{ruledtabular}
	\vspace*{2mm}
	\centering
	\begin{tabular}{*{5}{c}}
    	          & $T/T_c$ &$V_0$ & $2V_0$ & $3V_0$ \\
	\hline
	 $\frac{F}{VT^4}$  	& 0.970 & -0.7077(7) & -0.7782(6) & -0.8271(7) \\
	\vspace*{2mm}
	 		& 1.274 & -0.3713(10) & -0.3977(8) & -0.4227(9) \\
	 $\lan n_L\ran$	& 0.970 & 0.253(11) & 0.260(19)  & 0.289(35) \\
 	\vspace*{2mm}
	 		& 1.274 & 0.063(4) & 0.066(7) & 0.069(12) \\
	 $\lan n_M\ran$	& 0.970 & 0.253(11) & 0.264(19) & 0.259(31) \\
 	\vspace*{2mm}
	 		& 1.274 & 0.165(10) & 0.165(16) & 0.172(29) \\
	\end{tabular}
	\caption{Volume dependence of the free energy density and ensemble averages of dyon densities at several temperatures, for $V=V_0,2V_0\text{ and }3V_0$ (with $V_0=43.37$) respectively.}
	\label{Tab:Vol_Eff}
	\end{ruledtabular}
\end{table}

\begin{figure}[t]
\centering
\includegraphics[width=\textwidth]{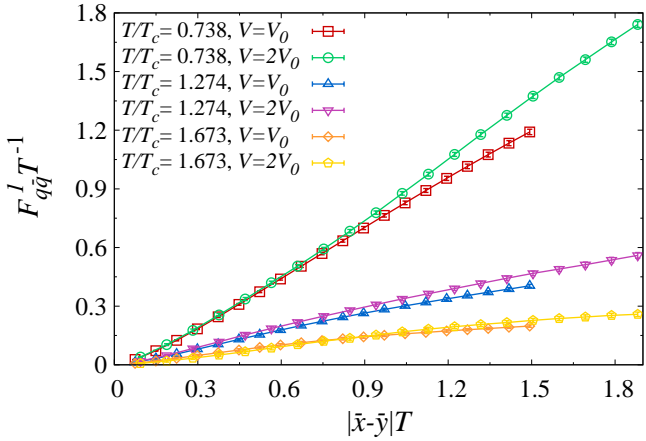}
\caption{Volume effects on the singlet static quark-anti-quark potential in both confined and deconfined phases.}
\label{Fig:FqqSing_Vol}
\end{figure}

The finite volume also bears influence on the evaluation of spatial correlation observables, in particular the static quark-anti-quark potential. As mentioned in the previous section, it exhibits unnatural behavior  near the largest distances that are allowed by the finite volume.  To test if this could indeed be a consequence of the finite volume, we've computed these observables with enlarged volume for comparison. In \cref{Fig:FqqSing_Vol} we show the results of the singlet channel potential calculated in a box twice the volume of the original volume $V_0$. For comparison, we also include the results from the original volume. One can see that indeed, the curved tails only appear at the edge of the box and at intermediate distances both potentials match substantially well. This comparison justifies our previous extraction of string tension via linear fit in intermediate distance regime and do suggest that for such spatial correlations, a significantly larger volume may be needed for their accurate evaluation. 

\vspace{0.5cm}

\subsection{The Influence of Dyon--anti-dyon Short-Range Correlation} \label{Sec:Core_Eff}

\begin{figure}[]
\centering
\includegraphics[width=1\textwidth]{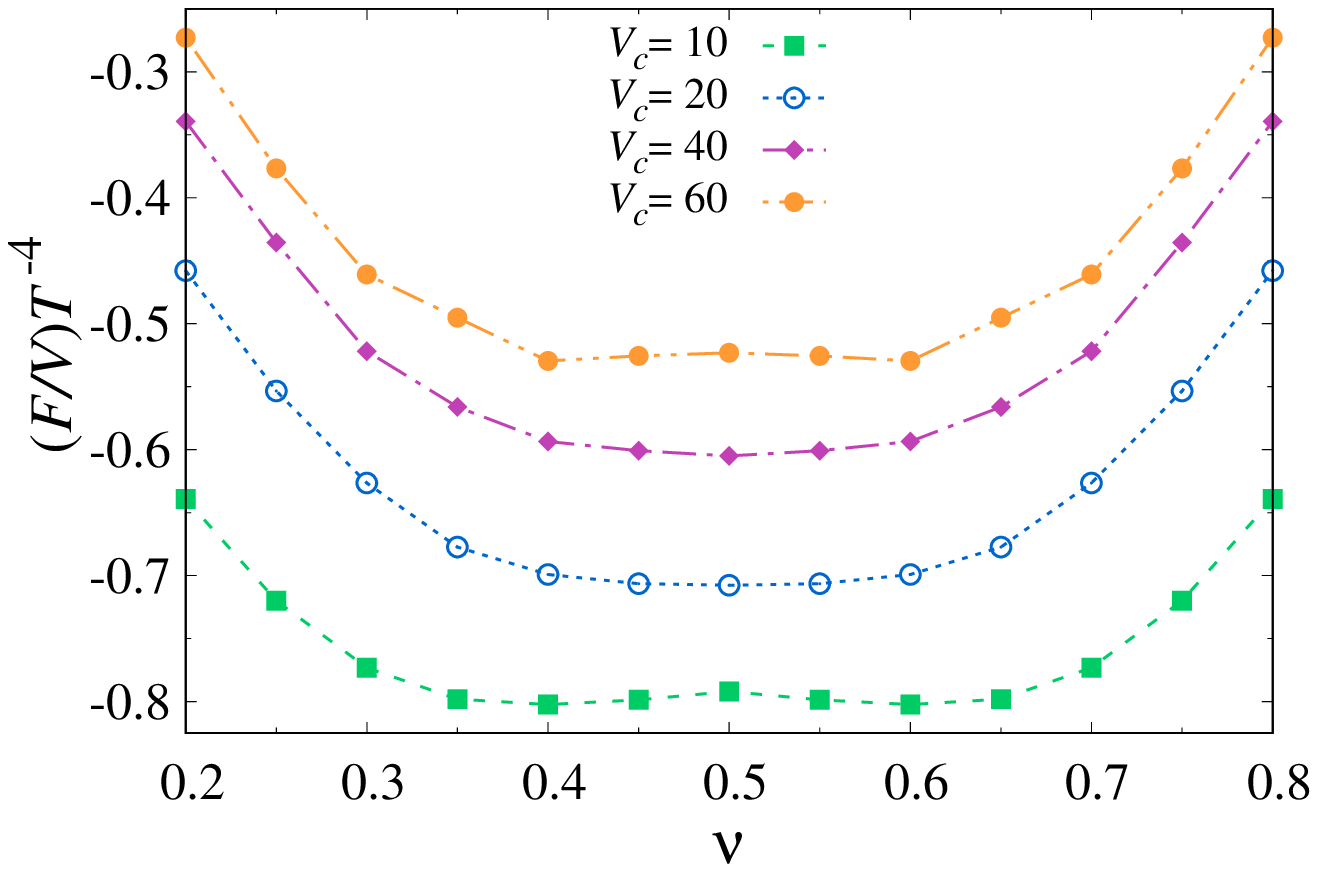}
\caption{The free energy density versus holonomy at $T/T_c=0.970$ $(S=7)$, for several different values of core potential strength $V_c$.}
\label{Fig:FDensity_V01}
\end{figure}

\begin{figure}[]
\centering
\includegraphics[width=1\textwidth]{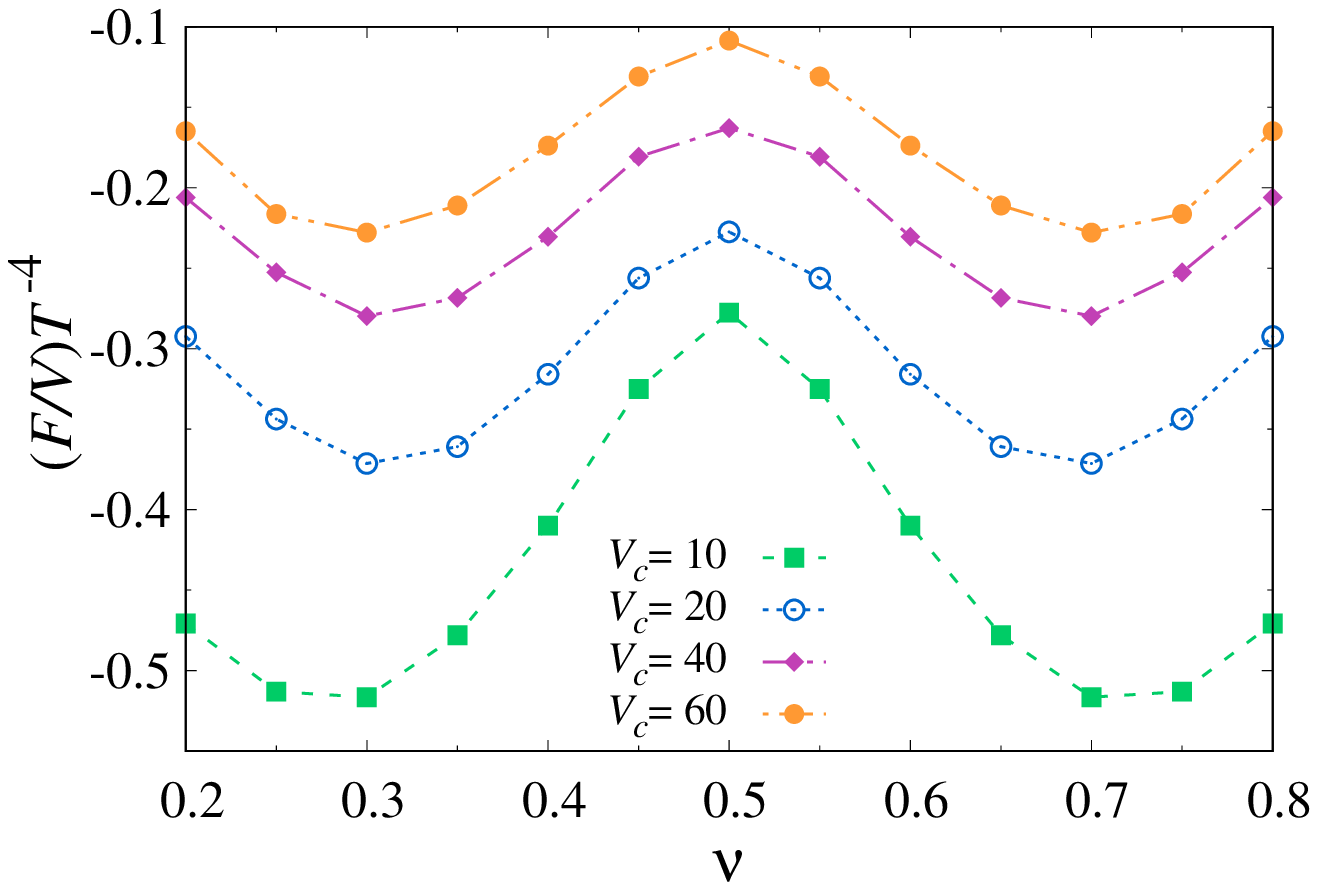}
\caption{The free energy density versus holonomy at $T/T_c=1.274$ $(S=9)$, for several different values of core potential strength $V_c$.}
\label{Fig:FDensity_V02}
\end{figure}

A key ingredient in  the confinement mechanism of dyon ensemble  is the repulsive core potential $V_{j\bar{j}}^C$. As defined in \cref{Eq:VDD_Core}, there are two parameters which quantify such interaction: $V_c$ is the strength and $\zeta_j^c$ the size of the core. It is important to understand the influence of these parameters on the various  observables. In \cref{Fig:FDensity_V01,Fig:FDensity_V02} we show the free energy density as a function of $\nu$ for different values of $V_c$ at both low and high temperatures. A general observation is that a larger core strength $V_c$ always favors more the confining holonomy $\nu=\bar{\nu}=1/2$. A smaller $V_c$, on the other hand, weakens the correlation and makes confinement harder to occur. Indeed for the $V_c=10$ case, even with the lowest temperature we explore, the system is still in the deconfined phase. These results also imply that the critical action $S_c$ needed for the confinement transition will shift toward larger values with increasing $V_c$.     

\begin{figure}[]
\centering
\includegraphics[width=1\textwidth]{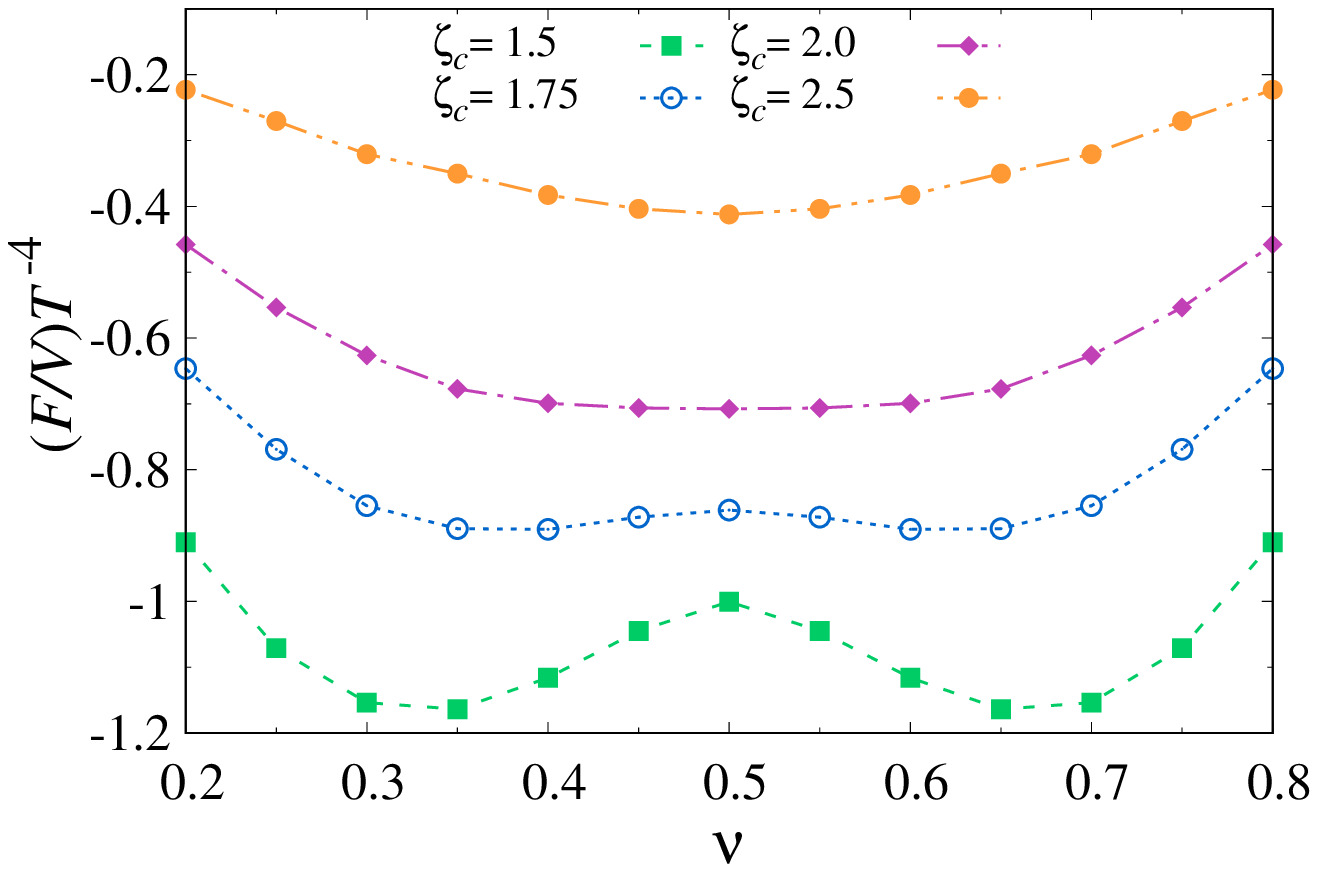}
\caption{The free energy density versus holonomy at $T/T_c=0.970$ $(S=7)$, for several different values of core potential range $\zeta^c_j$.}
\label{Fig:FDensity_x01}
\end{figure}

\begin{figure}[]
\centering
\includegraphics[width=1\textwidth]{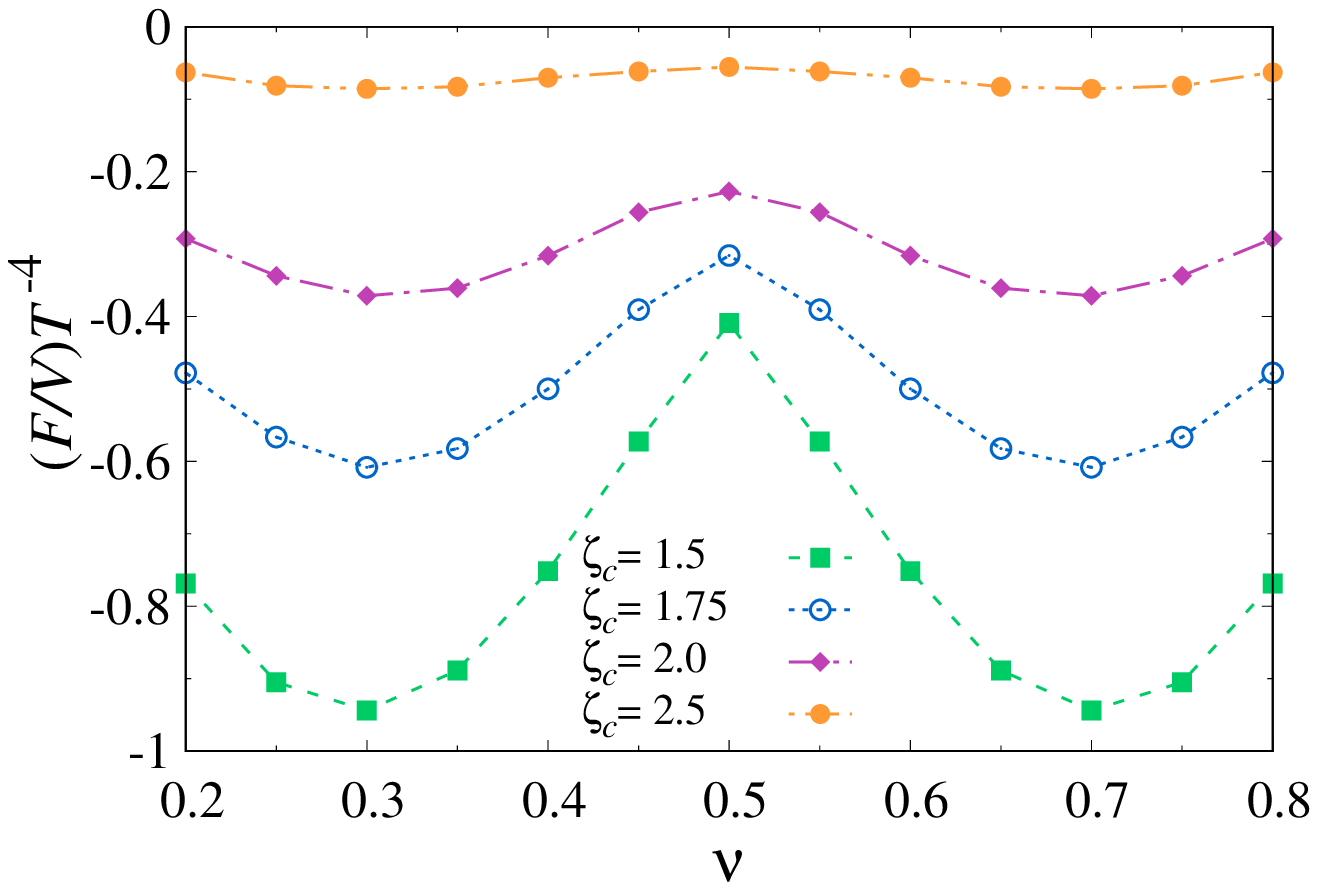}
\caption{The free energy density versus holonomy at $T/T_c=1.274$ $(S=9)$, for several different values of core potential range $\zeta^c_j$.}
\label{Fig:FDensity_x02}
\end{figure}

In a similar fashion, a change in the core size parameter $\zeta_j^c$, will also result in considerable effect on the behavior of the free energy density.

To investigate the influence of this parameter, we have computed the $\nu$ dependence of the free energy density for $\zeta_j^c=1.5,1.75$ and $2.5$ (in comparison with the standard choice of $\zeta_j^c=2$) with the results shown in \cref{Fig:FDensity_x01,Fig:FDensity_x02} for both low and high temperatures. (Due to the finite volume limitation, it is technically difficult to explore even larger core size values.) 

As can be seen, when the core size is decreased, the free energy density's minimum shifts further and further away from the confining holonomy value $\nu=\bar{\nu}=1/2$. If the core size is too small then the system would be in deconfined phase even at the low temperature value computed here. With a large core size, the system could maintain a holonomy value near the confining one even at high temperature.  The comparison clearly demonstrates the importance of the repulsive core. It is a strong repulsive core that drives the system toward favoring the confining holonomy at low temperature.

\subsection{The Debye Screening Mass} \label{Sec:MD_Eff}

\begin{figure}[hbt!]
\centering
\includegraphics[width=1\textwidth]{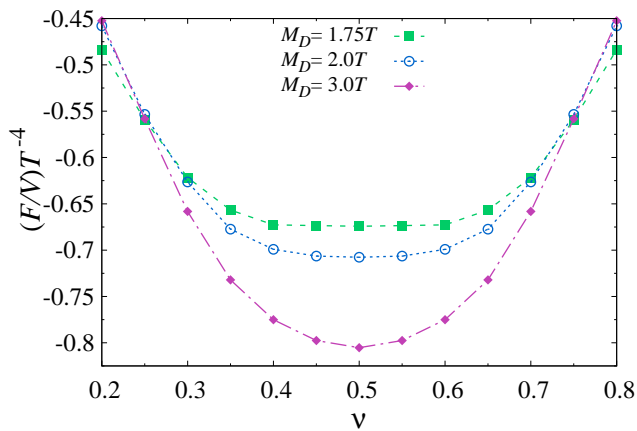}
\caption{The free energy density versus holonomy at $T/T_c=0.970$ $(S=7)$, for several different values of  Debye screening mass $M_D$.}
\label{Fig:FDensity_MD1}
\end{figure}

\begin{figure}[t]
\centering
\includegraphics[width=1\textwidth]{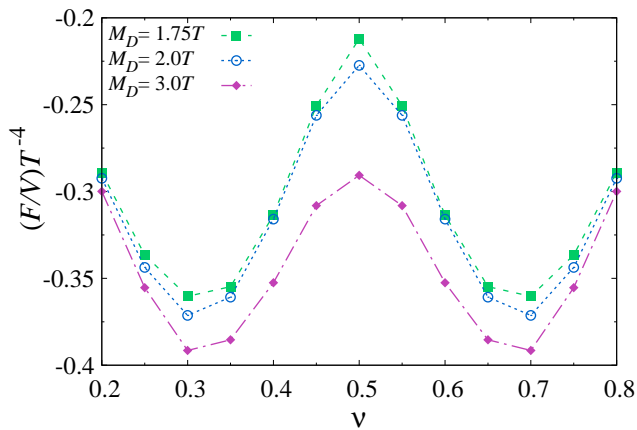}
\caption{The free energy density versus holonomy at $T/T_c=1.274$ $(S=9)$, for several different values of  Debye screening mass $M_D$.}
\label{Fig:FDensity_MD2}
\end{figure}

Finally, we investigate another important parameter for the ensemble, namely, the Debye mass $M_D$ used to regularize the large distance behavior of the Coulomb terms and therefore to account for the screening effect. This parameter plays an important role in controlling the contributions to the free energy from the long range Coulomb interactions among the dyons/anti-dyons. To see its effect, we compare the free energy density versus holonomy from dyon ensembles with three different choices of the $M_D$ in \cref{Fig:FDensity_MD1,Fig:FDensity_MD2} at both low and high temperatures. The results show that a smaller screening mass would disfavor the confining holonomy while a larger screening mass would help strengthen the confinement. This could be understood as follows: with a large screening mass the contribution to the free energy from many-body long-range Coulomb interactions get suppressed and thus the contribution from the short range correlations via the repulsive core, which essentially drives confinement,  become relatively more important.

\vspace{0.5cm}

\section{Conclusion}

Confinement is a remarkable nonperturbative phenomenon in pure Yang-Mills and QCD-like theories. The mechanism of confinement remains a significant challenge to our understanding and is generally believed to be a consequence of certain nontrivial topological configurations of the gluonic sector. The recently found KvBLL caloron solutions with nontrivial holonomy, consisting of constituent dyons/anti-dyons, have provided a concrete and promising path of investigation. In this paper, we have constructed a statistical ensemble of such correlated instanton-dyons and performed a thorough numerical study of its various properties for the $SU(2)$ Yang-Mills theory. Our main conclusion is that such an ensemble correctly produces the various essential features of the confinement dynamics from above to below the transition temperature. These features include the evolution of holonomy potential with temperature, a second order phase transition in terms of the order parameter (Polyakov loop expectation value), the linear static quark-anti-quark potential at large distance, etc. We have also found that the confinement dynamics is very sensitive to both the implemented short-range dyonanti-dyon correlations and the Debye screening effect in the many-body ensemble, by quantitatively investigating how the holonomy potential changes with these parameters. Given such success, it appears reasonable to believe that the ensemble of correlated instaton-dyons may indeed hold the key of confinement mechanism. The natural next steps of investigation would be the extension of the present framework toward the $SU(3)$ case as well as toward the inclusion of dynamic fermions thus allowing the study of nontrivial interplay between the confinement transition and the spontaneous chiral symmetry breaking, which we shall report elsewhere in the future.

\vspace{0.5cm}

\section*{Acknowledgements}
The authors are particularly grateful to E. Shuryak for many helpful discussions. The authors also thank E. Ilgenfritz, R. Larsen, R. Pisarski, I. Zahed, M. Faber and A. Zhitnitsky for useful discussions and communications. This work is supported by the National Science Foundation under Grant No. PHY-1352368. MALR is in addition supported by CONACyT under Doctoral supports Grants No. 669645. The computation of this research was performed on IU's Big Red II cluster that is supported in part by Lilly Endowment, Inc. (through its support for the Indiana University Pervasive Technology Institute) and in part by the Indiana METACyt Initiative.  

\vspace{0.5cm}

\begin{appendix}
\counterwithin{figure}{section}
\counterwithin{table}{section}


\section{Holonomy, Polyakov Loop, and Center Symmetry} \label{AP:Hlmy}

The classification of fiber bundles is an interesting topic in the geometry. Holonomy group, which describes the vector parallel transportation around closed loops, is one of tools to characterize the connection structure of a bundle. In gauge field theory the Wilson loop plays the same role as holonomy for gauge connections

\begin{flalign}
W[A_\mu]=\mathcal{P}\ {\rm exp}\left(i\oint \diff x^\mu A_\mu(x)\right).
\end{flalign}

In the imaginary-time formalism of finite temperature field theory the temporal direction is compactified to a circle of radius $(2\pi T)^{-1}$, where $T$ is the temperature. Therefore, the holonomy could be defined around this loop as

\begin{flalign}
L[A_\mu]=\mathcal{P}\ {\rm exp}\left(i\int_0^{1/T}\diff x_4\ A_4(\vec{x},x_4) \right),
\end{flalign}
which is the so-called Polyakov loop. Here $A_4$ is an element of Lie algebra $\mathfrak{su}(N)$. And it could have different forms in different representations. Practically the Polyakov loop is very useful in the study of phase transition at finite temperature. The ``condensate" of Polyakov loop serves as the order parameter of the confinement-deconfinement transition in the pure Yang-Mills theory.

To relate the fundamental quark confinement with the Polyakov loop it is intuitive to consider the free energy of a single static color charge, i.e., a quark with infinite mass \cite{Greensite:2011zz}

\begin{flalign}
e^{-F_q/T}&={\rm Tr}_q \,e^{-H/T}/{\rm Tr}\,e^{-H/T}\nonumber\\
\nb
&=\frac{\sum_n\langle\Psi_n(1\ \text{quark})|e^{-H/T}|\Psi_n(1\ \text{quark})\rangle}
{\sum_n\langle\Psi_n(0\ \text{quark})|e^{-H/T}|\Psi_n(0\ \text{quark})\rangle}.\\
\end{flalign}

From the view point of path integral formalism the numerator is just an infinitely heavy quark propagating from $(\vec{x},0)$ to $(\vec{x},1/T)$. Considering the kinetic part suppression due to the large mass the only contribution should come from the gauge field term which is equivalent to the Polyakov loop up to a constant

\begin{flalign}
e^{-F_q/T}\propto \langle \Tr\, L\rangle.
\end{flalign}
Clearly, the quark confinement, which corresponds to $F_q=+\infty$, will induce $\langle \Tr\,L\rangle=0$. Otherwise $\langle \Tr\,L\rangle=1$ if the quark is totally free. In this sense the Polyakov loop could be treated as a order parameter for the confinement deconfinement transition in the pure Yang-Mills theory. And it is also apparent that the intension of confinement is much more than the condensate of Polyakov loop. More information could be revealed by studying the topological details of it.

In the fundamental representation at spatial infinity up to a global transformation \cite{Diakonov:2007nv, Diakonov:2009jq}

\begin{flalign}
L={\rm diag}(e^{2\pi i\mu_1}, e^{2\pi i\mu_2}, ..., e^{2\pi i\mu_N}).
\end{flalign}
Generators of $SU(N)$ group are all traceless, so eigenvalues should satisfy

\begin{flalign}
\mu_1+\mu_2+...+\mu_N=0.
\end{flalign}
With a global transformation we could order them as

\begin{flalign}
\mu_1\leq\mu_2\leq ... \leq\mu_N\leq\mu_{N+1}\equiv\mu_1+1.
\end{flalign}
And in this paper the term holonomy will be used especially to call the set $\{\mu_m|m=1,2,...,N\}$. The holonomy is said to be trivial if $L$ is in the center group $Z_N$. Because $Z_N$ only has $N$ one-dimensional complex irreducible representations,
there are $N$ choices for the trivial holonomy

\begin{equation}
\nb
\mu_m = \left\lbrace\begin{array}{ll}
k/N -1 &\text{when } m\leq k\\
k/N &\text{when } m>k
\end{array}\right. ,
\end{equation}
where $k=1,..., N$. Without triviality constraint there would be lots of choices for the holonomy. A typical nontrivial one is the so-called ``maximally non-trivial" one

\begin{flalign}
\mu_m=-\frac{1}{2}-\frac{1}{2N}+\frac{m}{N}.
\end{flalign}
Obviously, this is an equidistant one which yields to ${\rm Tr}L=0$.  

Phase transitions often involve symmetry breaking or restoration. So it is for the confinement/deconfinement transition. With the definition of the Polyakov loop we can see that its condensate is the order parameter for the center symmetry. In finite temperature field theory the operation of the center symmetry is defined through the twisted gauge transformation, which satisfies the boundary condition along the imaginary-time dimension

\begin{flalign}
U(\vec{x}, x_4+T^{-1})=z\ U(\vec{x}, x_4),
\end{flalign}
where $z\in Z_N$ which is in the center of gauge group $SU(N)$. It could be checked that the Yang-Mills
lagrangian density is invariant under the center symmetry transformation. On the other hand, under such a gauge transformation $U(\vec{x},x_4)$, the Polyakov loop transforms as

\begin{flalign}
\nb
L[A_\mu^{U}] &=L[U A_\mu U^\dagger+\frac{i}{g}U\partial_\mu U^\dagger]\\
&=U(\vec{x}, T^{-1})L[A_\mu]U^\dagger(\vec{x}, 0).
\end{flalign}
Using the boundary condition the trace of Polyakov loop should transform as

\begin{flalign}
\Tr\, L[A_\mu^U]=z \Tr\, L[A_\mu].
\end{flalign}

In confined phase the $\langle \Tr\,L \rangle=0$ means the center symmetry is preserved. While in deconfined phase it becomes nonzero which means the symmetry breaking.

\section{The Perturbative Contribution to Holonomy Potential} \label{AP:HolPol}
Equilibrium state should be determined by the free energy

\begin{flalign}
Z(T)&=\sum_n\langle n|e^{-H/T}P_{\rm phys}|n\rangle\nonumber\\
&=\int \mathcal{D}A\ {\rm exp}^{-S[A_\mu]}\nonumber\\
&=e^{-F(T)/T}.
\end{flalign}

Once the background configuration has been chosen

\begin{flalign}
\vec{A} &= 0\nonumber\\
A_4 &= 2\pi T{\rm diag}(\mu_1, \mu_2,..., \mu_N),
\end{flalign}
we can always do the 1-loop perturbative calculation above this mean field. The essential part is to complete the integration and summation for dressed propagators of gauge fields \cite{Gross:1980br,Weiss:1980rj}.

\begin{flalign}
\nb
{\log}\,{\det}(-D^2)&=\sum^N_{j, k=1}{\log}\,{\det}\{-[\partial_\mu+2\pi T(\mu_j-\mu_k)\delta_{4\mu}]^2\}\\
&-{\log}\,{\det}(-\partial^2).
\end{flalign}

It could be seen that the longitudinal part of $\delta A_i$, which is an artifact of the gauge field, will cancel with the
ghost part $\delta A_4$ \cite{Weiss:1980rj}. Taking the group measure into account and performing the Matsubara summation,
the holonomy dependent part of the perturbative potential energy is related to the integration and summation as

\begin{flalign}
G&=\sum_{n=-\infty}^{+\infty}\int \Diff{3}p\ {\log}[(\omega_n+C)^2+p^2]\nonumber\\
&=\int \Diff{3}p\ {\log}(1-2{\cos}\beta C e^{-\beta p}+e^{-2\beta p})\nonumber\\
&=8\pi\int p^2\diff p\ {\rm Re}[{\log}(1-e^{i\beta C} e^{-\beta p})]\nonumber\\
&=\frac{8\pi}{\beta^3}({\rm Li_4}(e^{i\beta C})+{\rm Li_4}(e^{-i\beta C}))\nonumber\\
&=\frac{8\pi}{\beta^3}\frac{(2\pi)^4}{24}{\rm B_4}(\frac{(\beta C)_{{\rm mod}2\pi}}{2\pi})\nonumber\\
&=\frac{8\pi}{\beta^3}\frac{(2\pi)^4}{24}\left\{\frac{1}{16}\left[(2\frac{(\beta C)_{{\rm mod}2\pi}}{2\pi}-1)^2-1\right]^2-\frac{1}{30}\right\}\nonumber\\\nonumber
&=\frac{\pi^5 T^3}{3}\left\{\left[\left(\left(\frac{C}{\pi T}\right)_{{\rm mod}2}-1\right)^2-1\right]^2-\frac{8}{15}\right\},\\
\end{flalign}
where holonomy independent parts have been omitted at the second equation. And the ${\rm Li}_4(z)$ and ${\rm B}_4(z)$ are the polylogarithm function and Bernoulli polynomials of fourth order respectively. Here $C$ represents different combinations $\mu_j-\mu_k$ and $\beta=T^{-1}$.

Gathering all of contributions from different combinations $C=\mu_j-\mu_k$, up to a holonomy independent constant the perturbative potential energy for the $SU(N)$ case is obtained as

\begin{flalign}
\nb
P^{\text{pert}}&=V\frac{(2\pi)^2T^3}{3}\sum^N_{m>n}(\mu_m-\mu_n)^2[1-(\mu_m-\mu_n)]^2.\\
\label{Eq:PPert1}
\end{flalign}
For the $SU(2)$ case there is only one term $(\mu_2-\mu_1)^2[1-(\mu_2-\mu_1)]^2$. The maximally non-trivial holonomy is obtained from the above equation, leading to

\begin{flalign}
P^{\text{pert, max}} &= V\frac{(2\pi)^2T^3}{3}\sum_{i=1}^{N-1}\frac{i^2}{N^2}(1-\frac{i}{N})^2\nonumber\\
&=V\frac{(2\pi)^2T^3}{180}\frac{N^4-1}{N^2},
\end{flalign}
where the Faulhaber's formula are used to complete the summation $\sum_{i=1}^N i^m$. At this 1-loop level the perturbative potential energy has $N$ minima corresponding to $N$ elements of the center group. And the confining holonomy gives larger potential energy than the trivial ones. This means at 1-loop level trivial holonomies, which indicate the deconfinement, are favored
at arbitrary temperatures. Hence, in order to achieve confinement at low temperature a more strict calculation is necessary with a topological non-trivial background configuration. The KvBLL Caloron is one of these choices.

\section{The KvBLL Caloron Solution} \label{AP:KvBLL}

The caloron field with non-trivial holonomy discovered by Kraan and van Baal \cite{Kraan:1998pm, Kraan:1998sn} and independently by Lee and Lu \cite{Lee:1998bb} (therefore also known in the literature as the \textit{KvBLL caloron}), is a classical solution to the $SU(N)$ Yang-Mills equations of motion in $\mathbb{R}^3\times S^1$. It is a self-dual field with unit topological charge and most importantly, the $A_4$ component can be gauged to be diagonal and constant at spatial infinity, which leads to a non-trivial Polyakov loop.

In the periodic gauge, the $SU(2)$ KvBLL caloron field with period $1/T$ is given by

\begin{flalign}
\nb
A_\mu^{\text{KvBLL}} &=  \delta_{\mu4} v\frac{\tau^3}{2} + \frac{\tau^3}{2}\bar{\eta}^{3}_{\mu\nu}\pt_\nu\log\Phi \\
\nb
& + \frac{\Phi}{2}\Re\lsb \lb\bar{\eta}_{\mu\nu}^1 -i\bar{\eta}_{\mu\nu}^2\rb \lb\tau^1 + i\tau^2\rb \right.\\
& \times \left.\lb \pt_\nu + iv\delta_{\nu4}\rb\tilde{\chi}\rsb,
\label{Eq:KvBLL_Solution}
\end{flalign}
where 

\begin{flalign}
\nb
\hat{\psi} &=  -\cos(2\pi Tx_4) + \cosh(\bar{v}r)\cosh(vs) \\
\nb
&+ \frac{r^2 +s^2 -\pi^2\rho^4 T^2}{2rs}\sinh(\bar{v}r)\sinh(vs), \\
\nb
\psi &= \hat{\psi} + \frac{\pi^2\rho^4T^2}{rs}\sinh(\bar{v}r)\sinh(vs) \\
\nb
& +\pi\rho^2T\lsb\frac{\sinh(vs)\cosh(\bar{v}r)}{s} + \frac{\sinh(\bar{v}r)\cosh(vs)}{r}\rsb,\\
\nb
\tilde{\chi} &= \frac{\pi\rho^2T}{\psi}\lsb e^{-2\pi ix_4} \frac{\sinh(vs)}{s} + \frac{\sinh(\bar{v}r)}{r}\rsb, \\
\Phi &=\frac{\psi}{\hat{\psi}}.
\end{flalign}
Here, $\bar{\eta}_{\mu\nu}^a\equiv \varepsilon_{\ mu\nu}^a -\delta_\mu^a\delta_{\nu4}+\delta_\nu^a\delta_{\mu4}$ are the so called 't Hooft symbols, $T$ the temperature and $\tau^a$ the Pauli matrices. The meaning of the $s$ and $r$ variables will be explained shortly. From this expression, it is not hard to see that at spatial infinity, the fourth component is indeed diagonal and constant $A_4|_{|\vec{x}|\to\infty}=v\frac{\tau^3}{2}$. This asymptotic value is parametrized as $v\equiv 2\pi T\nu$, with $\nu\in[0,1]$ and analogously, $\bar{v}=2\pi T\bar{\nu}$ with $\bar{\nu}=1-\nu$. Thus, the trace of the Polyakov loop at spatial infinity has the non-trivial form

\begin{flalign}
\nb
L_{\infty} &\equiv  \lim_{|\vec{x}|\to\infty}\frac{1}{2}\Tr\,\mathcal{P}\exp\lb i\int_0^{1/T}\diff x_4 A_4^{\text{KvBLL}}\rb \\
 &= \cos(\pi\nu),
\label{Eq:PolInfty}
\end{flalign}
where $\nu=\frac{1}{2}$ corresponds to maximal non-trivial holonomy $(L_\infty=0)$ and $\nu=0$ trivial holonomy $L_\infty=1)$. Therefore, $\nu$ is naturally called the \textit{holonomy parameter}.

The anti-self-dual caloron or \textit{anticaloron} $\bar{A}_\mu$ with negative topological charge is easily obtained from \cref{Eq:KvBLL_Solution} by

\begin{flalign}
\nb
\bar{A}_4^{\text{KvBLL}}(\vec{x},x_4)&=A_4^{\text{KvBLL}}(-\vec{x},x_4), \\
\bar{A}_i^{\text{KvBLL}}(\vec{x},x_4)&=-A_i^{\text{KvBLL}}(-\vec{x},x_4).
\end{flalign}

As expected, the KvBLL reduces to the Harrington-Shepard caloron \cite{Harrington:1978ve} in the limit of trivial holonomy $(\nu\to0$ or $\bar{\nu}\to0)$. Furthermore, it becomes a standard BPST instanton \cite{Belavin:1975fg} of size $\rho$ in the zero temperature limit.

\begin{table}
	{\caption{Properties of the $SU(2)$ (anti)dyons.}\label{Tab:Dyon_Prop}}
	\begin{ruledtabular}
	\begin{tabular}{*{5}{c}}
    & $M$ & $\bar{M}$ & $L$ & $\bar{L}$ \\
	\hline
	$\begin{array}{c}
	\text{Electric}\\ \text{charge}
	\end{array}$ & 1 & 1 & -1 & -1 \\
	$\begin{array}{c}
	\text{Magnetic}\\ \text{charge}
	\end{array}$ & 1 & -1 & -1 & 1 \\
	\vspace*{2mm}
	Action & $\nu\frac{8\pi^2}{g^2}$ & $\nu\frac{8\pi^2}{g^2}$ & $\bar{\nu}\frac{8\pi^2}{g^2}$ & $\bar{\nu}\frac{8\pi^2}{g^2}$ \\
	Radius & $v^{-1}$ & $v^{-1}$ & $\bar{v}^{-1}$ & $\bar{v}^{-1}$\\
	\end{tabular}
	\end{ruledtabular}
\end{table}

One of the most important properties of this solution becomes relevant when $\rho\gg1/T$. In this limit, the field is seen as composed of two constituent monopoles separated by a distance $\pi\rho^2 T$. As $\rho\to\infty$, the caloron becomes static and the monopoles are identified as the BPS type \cite{Bogomolny:1975de, Prasad:1975kr} with unit, but opposite, electric and magnetic charges, therefore named \textit{dyons} or in this context \textit{intanton-dyons}.  

(Anti)dyons are commonly known as (anti)self-dual static solutions of the Yang-Mills equations of motion with an adjoint scalar (Higgs) field. However, one can construct dyonic solutions in pure Yang-Mills theory with the condition of non-trivial holonomy, namely $\left.A_4\right|_{|\vec{x}|\to\infty}=v$. For $SU(2)$ there are four kinds of dyon solutions which following the usual convention in the literature are labeled $M$ and $L$ for the self-dual fields and $\bar{M}$ and $\bar{L}$ for the antiself-dual ones, referred to as \textit{anti-dyons} (--- see \cref{Tab:Dyon_Prop}). In the \textit{hedgehog} gauge, the $M$ fields have the form of the common BPS monopole solution(for more details on the derivation refer to \cite{Diakonov:2009jq,Diakonov:2002qw})

\begin{flalign}
\nb
\vspace*{2mm}
A_4^{M,\bar{M}} &= \mp n_a \lb v\coth(v|\vec{x}|) - \frac{1}{|\vec{x}|} \rb\frac{\tau^a}{2}, \\
A_i^{M,\bar{M}} &= \varepsilon_{aij}n_j \lb \frac{1}{|\vec{x}|} - \frac{v}{\sinh(v|\vec{x}|)} \rb\frac{\tau^a}{2},
\label{Eq:Dyon_Hdg}
\end{flalign}
where $n_a=x_a/|\vec{x}|$ and the (lower)upper sign corresponds to the (anti)self-dual solution. 

If a gauge configuration consists of more than two dyons, it is inconvenient to superimpose them in this gauge, since we are interested in configurations where all dyons have the same $A_4$ asymptotics at spatial infinity. This is achieved by using the matrices 

\begin{flalign}
\nb
S_+ &= e^{-i\frac{\phi}{2}\tau^3}e^{i\frac{\theta}{2}\tau^2}e^{i\frac{\phi}{2}\tau^3}, \\
\nb
 S_- &= -i\tau^2S_+ \\
 &= e^{i\frac{\phi}{2}\tau^3}e^{i\frac{\theta-\pi}{2}\tau^2}e^{i\frac{\phi}{2}\tau^3},
\end{flalign}
which satisfy the identity $S_\pm(n_a \tau^a)S_\pm^\dagger = \pm \tau^3$, and gauge-transform the dyon fields \cref{Eq:Dyon_Hdg} as

\begin{equation}
A_\mu^{M,\bar{M}} \rightarrow S_\mp  A_\mu^{M,\bar{M}} S_\mp^\dagger + i S_\mp \pt_\mu S_\mp^\dagger.
\end{equation}

In spherical coordinates, the dyon solutions in the new gauge take the form

\begin{flalign}
\nb
A_4^{M,\bar{M}} &=  \frac{\tau^3}{2}\lb v\coth(v|\vec{x}|) - \frac{1}{|\vec{x}|}\rb ,\\
\nb
\pm A_i^{M,\bar{M}}&= \left\lbrace \begin{array}{lll}
\vspace*{2mm}
A_r &= 0,\\
\vspace*{2mm}
A_\theta &= \dfrac{v}{2\sinh(v|\vec{x}|)}\left(\tau^1\sin\phi + \tau^2\cos\phi \right),\\
\vspace*{2mm}
A_\phi &= \dfrac{v}{2\sinh(v|\vec{x}|)}\left(\tau^1\cos\phi - \tau^2\sin\phi \right) \\
& + \dfrac{\tau^3}{2}\dfrac{\tan\frac{\theta}{2}}{|\vec{x}|}.
\end{array} \right.\\
\label{Eq:M_SG}
\end{flalign}

One should notice first that now the $A_4$ component is Abelian and equal for both $M$ and $\bar{M}$. Moreover, we have introduced a singularity along the negative $x_3$-axis in $A_\phi$, a so called Dirac string which is merely a consequence of the gauge choice, hence the name \textit{stringy gauge}.

The $L$ and $\bar{L}$ solutions are obtained from \cref{Eq:M_SG} by replacing $v\to\bar{v}$ and apply two gauge transformations: first the time dependent $U_1=\exp(-i\pi Tx_4\tau^3)$ followed by a global rotation $U_2=\exp(i\pi\tau^2/2)$ \cite{Lee:1998bb,Lee:1997vp,Lee:1998vu}. As required, these will leave the asymptotics of $A_4$ in the same form as for the $M$ type solutions with the caveat that the spatial components are no longer static; however, in the large distance limit, neglecting exponentially small terms, the time dependent terms vanish and are no longer relevant in the scope of this article. The $L$ type dyon fields in the stringy gauge thus are

\begin{flalign}
\nb
A_4^{L,\bar{L}} &=  \frac{\tau^3}{2}\lb 2\pi T - \bar{v}\coth(\bar{v}|\vec{x}|) + \frac{1}{|\vec{x}|}\rb ,\\
\nb
\pm A_i^{L,\bar{L}}&=\left\lbrace \begin{array}{lll}
\vspace*{2mm}
A_r &= 0,\\
\vspace*{2mm}
A_\theta &= \dfrac{\bar{v}}{2\sinh(\bar{v}|\vec{x}|)}\lsb\tau^1\sin(2\pi Tx_4-\phi)\right. \\
\vspace*{2mm}
& + \left. \tau^2\cos(2\pi Tx_4-\phi)\rsb,\\
\vspace*{2mm}
A_\phi &= \dfrac{\bar{v}}{2\sinh(\bar{v}|\vec{x}|)}\lsb -\tau^1\cos(2\pi Tx_4-\phi) \right. \\
& + \left. \tau^2\sin(2\pi Tx_4-\phi) \rsb - \dfrac{\tau^3}{2}\dfrac{\tan\frac{\theta}{2}}{|\vec{x}|}.
\end{array} \right.\\
\label{Eq:L_SG}
\end{flalign}

\begin{figure}[t]
\centering
\includegraphics[width=\textwidth]{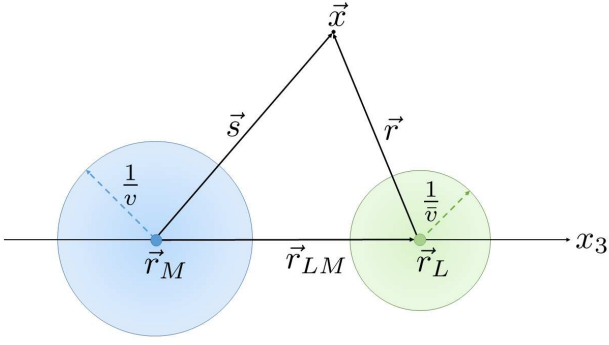}
\caption{Coordinates of the KvBLL caloron in terms of the center of mass positions of its constituent dyon fields.}
\label{Fig:KvBLL_Coordinates}
\end{figure}

Going back to the KvBLL field, the emergence of such configurations suggests to express the caloron in terms of the ``constituent" dyon's positions.  The coordinates used to write the caloron in \cref{Eq:KvBLL_Solution} are then the positions of the dyon's center of mass denoted by $\vec{r}_L$ and $\vec{r}_M$, the dyon separation $r_{LM}\equiv \left|\vec{r}_L-\vec{r}_M\right|=\pi\rho^2 T$, which for convenience is chosen to be along the $x_3$-axis (--- see  \cref{Fig:KvBLL_Coordinates}); i.e. $\vec{r}_{LM}=r_{LM}\hat{e}_3$, and the distances from the observation point $\vec{x}$ to the dyon centers: $\vec{s}=\vec{x}-\vec{r}_M$ and $\vec{r}=\vec{x}-\vec{r}_L$. 

This monopole picture is more evident when looking at the caloron in the vicinity of one of its constituent dyons and far away from the other, namely at large separations. For instance, near the $L$ dyon center and far away from the $M$ dyon $(s\gg1/v)$, the caloron field reduces to that of the $L$ dyon, whose asymptotic behavior is given by (--- see \cref{Eq:L_SG})

\begin{flalign}
\nb
A_4^{L,\bar{L}}|_{r\to\infty} &= \frac{\tau^3}{2}\lb v + \frac{1}{r} \rb, \\
 A_\phi^{L,\bar{L}}|_{r\to\infty} &= \mp\frac{\tau^3}{2} \frac{\tan\frac{\theta}{2}}{r}, 
\end{flalign}
where $\phi$ and $\theta$ are the polar and azimuthal angles in spherical coordinates centered at $\vec{r}_L$. The other components vanish in this limit. Analogously, near the $M$ dyon and far away from the $L$ $(r\gg1/\bar{v})$, the field is that of the $M$ dyon with asymptotics (--- see \cref{Eq:M_SG})

\begin{flalign}
\nb
A_4^{M,\bar{M}}|_{s\to\infty} &= \frac{\tau^3}{2}\lb v - \frac{1}{s} \rb, \\
 A_\phi^{M,\bar{M}}|_{r\to\infty} &= \pm\frac{\tau^3}{2} \frac{\tan\frac{\theta}{2}}{s}. 
\end{flalign}

Finally, in the limits $s\gg1/v$ and $r\gg1/\bar{v}$, but not necessarily at large separations $r_{LM}$; the KvBLL caloron field also becomes Abelian and takes the form

\begin{equation}
A_\mu^{\text{KvBLL}}=\frac{\tau^3}{2}\lb\delta_{\mu4}v + \bar{\eta}^3_{\mu\nu}\pt_\nu\log\Phi \rb,
\label{Eq:KvBLL_Asym}
\end{equation}
where $\Phi$ in this limit reduces to

\begin{equation}
\Phi=\frac{r+s+r_{LM}}{r+s-r_{LM}}.
\end{equation}

The only nonvanishing components of \cref{Eq:KvBLL_Asym} are

\begin{flalign}
\nb
A_4^{\text{KvBLL}} &= \frac{\tau^3}{2}\lb v + \frac{1}{r} -\frac{1}{s} \rb,\\
\nb
A_\phi^{\text{KvBLL}} &= -\frac{\tau^3}{2}\lb \frac{1}{r} +\frac{1}{s} \rb \\
\nb
& \times\sqrt{\frac{(r_{LM}-r+s)(r_{LM}+r-s)}{(r_{LM}+r+s)(r+s-r_{LM})}}.\\
\end{flalign}

\section{The Quantum Weight} \label{AP:KvBLL_QW}

In a similar fashion as it was done for the BPST instanton (at $T=0$) \cite{'tHooft:1976fv} and for the Harrington-Shepard caloron \cite{Gross:1980br} (at $T\neq0$), it is of interest to calculate the contribution of small quantum oscillations of the KvBLL caloron to the Yang-Mills partition function 

\begin{flalign}
\nb
\mathcal{Z}&= \int\mathcal{D}A_\mu \,e^{-S[A_\mu]}\\
 &= \int\mathcal{D}A_\mu\exp\lsb-\frac{1}{2g^2}\int \Diff4x \,\Tr\, F_{\mu\nu}F_{\mu\nu}\rsb.
\end{flalign}

In broad terms, this semiclassical procedure consists in taking the classical solution as a background field such that the gauge fields in the functional integral are

\begin{equation}
A_\mu(x)=A_\mu^{\text{KvBLL}}(x) + a(x),
\end{equation}
where $a(x)$ is a small quantum fluctuation of the classical solution (the KvBLL field). Then expand the action around the saddle point up to the desired order in $a_\mu$ and compute the functional integral.

In \cite{Diakonov:2004jn} Diakonov et al. obtained an analytic expression for the quantum weight of the $SU(2)$ KvBLL caloron in the one-loop approximation. They showed that in the limit of large separation between the constituent dyons (in the temperature scale) $r_{LM}\gg 1/T$, it can be written as 

\begin{flalign}
\nb
\mathcal{Z}_{\text{KvBLL}} &= e^{-VP(\nu)/T}\int \Diff3r_L\Diff3r_M T^6 C 2\pi \lb\frac{8\pi^2}{g^2}\rb^4 \\
\nb
& \times \lb\frac{\Lambda_{\text{PV}}\, e^{\gamma_E}}{4\pi T}\rb^{\frac{22}{3}} \lb\frac{1}{Tr_{LM}}\rb^{\frac{5}{3}} \lb 1+2\pi T\nu\bar{\nu}r_{LM}\rb \\
\nb
&\times \lb 1 + 2\pi T\nu r_{LM} \rb^{\frac{8\nu}{3}-1}\lb 1 + 2\pi T\bar{\nu} r_{LM} \rb^{\frac{8\bar{\nu}}{3}-1},\\
\label{Eq:Z_KvBLL1}
\end{flalign} 
where $P(\nu)=(4\pi^2/3)T^4\nu^2\bar{\nu}^2$ is the one-loop perturbative potential \cite{Gross:1980br, Weiss:1980rj} (--- see \cref{AP:HolPol}), $C\approx 1.03142$ is a combination of universal constants and the linear term in $r_{LM}$ proportional to $P''(\nu)$ from the exponential factor has been ignored in this work.

This expression can be further simplified in the approximation where the separation between dyons is much larger than their core sizes $r_{LM}\gg \frac{1}{2\pi T\nu},\frac{1}{2\pi T\bar{\nu}}$; taking the form

\begin{flalign}
\nb
\mathcal{Z}_{\text{KvBLL}} &= e^{-VP(\nu)/T}\int\Diff3r_L\Diff3r_M T^6(2\pi)^{\frac{8}{3}}C\lb\frac{8\pi^2}{g^2}\rb^4\\
& \times \lb\frac{\Lambda_{\text{PV}}\, e^{\gamma_E}}{4\pi T}\rb^{\frac{22}{3}} \nu^{\frac{8}{3}\nu}\bar{\nu}^{\frac{8}{3}\bar{\nu}}.
\label{Eq:Z_KvBLL2}
\end{flalign}

To obtain \cref{Eq:Z_KvBLL1}, one has to calculate the invariant measure of the moduli space metric of the caloron field denoted as $\sqrt{\det (g)}$. In the general case of $SU(N)$, this is shown to be exactly equal to the determinant of a $N\times N$ matrix $\hat{G}$ \cite{Diakonov:2005qa, Diakonov:2007nv}, which for $SU(2)$ is given by 

\begin{equation}
\sqrt{\det (g)}=\det (\hat{G}),
\label{Eq:G_KvBLL}
\end{equation}
where

\begin{equation}
\hat{G}=\begin{pmatrix}
\vspace*{2mm}
4\pi\bar{\nu}+\dfrac{1}{Tr_{LM}} & -\dfrac{1}{Tr_{LM}}\\
-\dfrac{1}{Tr_{LM}} & 4\pi\nu+\dfrac{1}{Tr_{LM}},
\end{pmatrix},\\
\end{equation}
which in the limit of large dyon separation reduces to $\det(\hat{G})\approx16\pi^2\nu\bar{\nu}$, and thus the partition function \cref{Eq:Z_KvBLL2} is rewritten as

\begin{flalign}
\nb
\mathcal{Z}_{\text{KvBLL}}& = e^{-VP(\nu)/T}\int\Diff3r_L\Diff3r_M T^6\frac{(2\pi)^{\frac{2}{3}}}{4}C\det (\hat{G})\\
\nb
& \times \lb\frac{8\pi^2}{g^2}\rb^4 \lb\frac{\Lambda_{\text{PV}}\, e^{\gamma_E}}{4\pi T}\rb^{\frac{22}{3}}\nu^{\frac{8}{3}\nu-1}\bar{\nu}^{\frac{8}{3}\bar{\nu}-1}.\\
\label{Eq:Z_KvBLL3}
\end{flalign}

The factor $\lb\frac{\Lambda_{\text{PV}}\, e^{\gamma_E}}{4\pi T}\rb^{\frac{22}{3}}$, appears from the running of the coupling constant $g$, in the Pauli-Villars regularization scheme. Namely

\begin{equation}
\lb\frac{\Lambda}{T}\rb^{\frac{22}{3}} = e^{-\frac{8\pi^2}{g^2(T)}},
\end{equation}
where we have absorbed all constants into $\Lambda$. At the one loop calculation, the $g^{-8}$ coupling in \cref{Eq:Z_KvBLL3} is not renormalized; however, a two loop improvement (ignoring the effects on $P(\nu)$) will give 

\begin{equation}
\lb\frac{8\pi^2}{g^2}\rb^4 \lb\frac{\Lambda}{T}\rb^{\frac{22}{3}} \to \lb\frac{8\pi^2}{g^2(T)}\rb^4 e^{-\frac{8\pi^2}{g^2(T)}}h(T/\Lambda), 
\label{Eq:2Loop}
\end{equation}
where

\begin{flalign}
\nb
h(T/\Lambda)&=\exp \left\lbrace -\frac{34}{11}\log\lsb 2\log\lb\frac{T}{\Lambda}\rb \rsb\right. \\
&  + \left.\frac{510}{1331}\frac{\log\lsb\frac{22}{3}\log\lb\frac{T}{\Lambda}\rb\rsb}{\log\lb\frac{T}{\Lambda}\rb} \right\rbrace. 
\end{flalign}

As an approximation, one can include the two loop improvement by substituting \cref{Eq:2Loop} and absorb the rest of the constant factors into a parameter $\Gamma$ which is modulated in the simulation and fixed to be $\Gamma\approx 0.119$.

Finally, the caloron quantum weight takes the form

\begin{flalign}
\nb
\mathcal{Z}_{\text{KvBLL}} &= e^{-VP(\nu)/T}\int\Diff3r_L\Diff3r_M \det (\hat{G})T^6\Gamma^2 \\
\nb
& \times S^4 e^{-S} \nu^{\frac{8}{3}\nu-1}\bar{\nu}^{\frac{8}{3}\bar{\nu}-1}\\
\nb
&=e^{-VP(\nu)/T}\int \lb\Diff{3}r_L \,f_L \rb \lb\Diff{3}r_M\, f_M \rb T^6 \det(\hat{G}),\\
\label{Eq:Z_KvBLL4}
\end{flalign}
with

\begin{equation}
f_M=\Gamma S^2 e^{-\nu S}\nu^{\frac{8\nu}{3}-1}, \quad f_L=\Gamma S^2 e^{-\bar{\nu} S}\bar{\nu}^{\frac{8\bar{\nu}}{3}-1}
\end{equation}
the respective dyon fugacities and the instanton action

\begin{equation}
 S(T)=\frac{8\pi^2}{g^2(T)}=\frac{22}{3}\log\lb\frac{T}{\Lambda}\rb.
 \end{equation}

\end{appendix}

\end{document}